\documentclass[epj]{svjour}
\usepackage{epsfig,pstricks}
\begin{document}
\title{Effects of the liquid-gas phase transition 
  and cluster formation on the symmetry energy}
\author{S. Typel\inst{1} \and 
H.H. Wolter\inst{2} \and
G. R\"{o}pke\inst{3} \and
D. Blaschke\inst{4,5}
}                     
%
%
\institute{GSI Helmholtzzentrum f\"{u}r Schwerionenforschung,
  Planckstra\ss{}e~1, 64291 Darmstadt, Germany \and
  Ludwig-Maximilians-Universit\"{a}t M\"{u}nchen, Am Coulombwall~1,
  85748 Garching, Germany \and
  Universit\"{a}t Rostock, Institut f\"{u}r Physik, 18051 Rostock,
  Germany \and
  Instytut Fizyki Teoretycznej, Uniwersytet Wroc\l{}awski, 
  pl. M. Borna 9, 50-204 Wroc\l{}aw, Poland \and
  Bogoliubov Laboratory for Theoretical Physics, JINR Dubna, 
  Joliot-Curie Str. 6, 141980 Dubna, Russia
}
\date{Received: date / Revised version: date}

%
\abstract{
Various definitions of the symmetry energy are introduced for nuclei,
dilute nuclear matter below saturation density
and stellar matter, which it found in compact stars or core-collapse 
supernovae. The resulting differences are
exemplified by calculations in a theoretical approach based on a
generalized relativistic density functional for dense matter.
It contains nucleonic clusters as explicit degrees of freedom with medium
dependent properties that are derived for light clusters from a
quantum statistical approach. With such a model the dissolution of
clusters at high densities can be described. The effects of the
liquid-gas phase transition in nuclear matter and of cluster
formation in stellar matter on the density dependence of the symmetry
energy are studied for different temperatures. It is observed that 
correlations and the formation of inhomogeneous matter 
at low densities and temperatures
causes an increase of the symmetry energy as compared to calculations
assuming a uniform uncorrelated 
spatial distribution of constituent baryons and leptons.
} 
\maketitle
\section{Introduction}
\label{intro}


The isospin degree of freedom \cite{Hei32,Wig37} is a particular feature of systems that
contain strongly interacting particles. The (a)symmetry energy
characterizes how much the energy of the system changes when the
isospin resp.\ the asymmetry is varied keeping other quantities
and quantum numbers constant. Historically, the symmetry energy appeared
first in the description of binding energies of finite
nuclei \cite{Wei35,Bet36a}. Later, the concept was generalized to
nuclear matter.
It has proven to be enormously valuable in nuclear physics.
The dependence of the symmetry energy on 
density and temperature is of particular interest because there are
tight connections to observable properties of atomic nuclei, neutron matter,
heavy-ion collisions and compact stars. Many studies are devoted to
understand and quantify these relations, see, e.g., the recent
works \cite{Agr13,Lat12,LiH13,Sot13,Wan13,Heb13,Zha13a,Don13,Che12,%
Mar12,Fat13,Rus12,Fat12,Don12,Gan12,Tra12,Tsa12}, 
references cited therein,
and the contributions to this volume of the European Physical Journal A.

A meaningful comparison of the symmetry energy in experiment and theory relies on a
precise definition of this quantity. Since different methods are used
to extract the symmetry energy from measurements and calculations, the
relation between the obtained values has to be understood. 
The deduced symmetry energies may depend on the particular system that is
investigated and thermodynamic conditions have to be
taken into account. These differences due to different definitions are more
pronounced at densities below the nuclear saturation density. Thus, we
will concentrate on this region in the present work. 

Dense matter will
be treated only in the thermodynamic limit of infinite volume and particle
number. Finite size systems with a limited number of particles, as they appear in
laboratory experiments, may differ in their properties from the
infinite systems. Their theoretical description requires to choose the appropriate
thermodynamic ensemble and the derived symmetry energies can depend
on this choice. 

In the laboratory, the equation of state can be studied in heavy-ion
collisions. 
In such processes the matter may even be out of thermodynamic equilibrium. The derivation of
densities and temperatures of equivalent systems in thermodynamic equilibrium is
rather involved and may depend on particular model assumptions.
However, this is not the topic of our present work.
A recent discussion of results from laboratory experiments with
heavy-ion collisions is given by Hagel et.\ al.\ \cite{Hag13} in this
volume. Recent experimental approaches to extract the nuclear symmetry
energy from light fragment yields are based on the isoscaling
method \cite{Kow07,Nat10,Wad12}. There the symmetry free energy is obtained by
studying collisions of nuclei with different neutron numbers as a
function of the asymmetry.


Dense matter is often treated as a spatially homogenous system of 
strongly interacting particles considering only baryons and mesons 
and the symmetry energy is calculated under this
condition in theoretical models. 
However, due to the interaction between the constituents,
correlations are an important feature that can lead to the formation
of clusters \cite{Rad09,Typ10,Bot10,Hem10,Hem11,Vos12,Roe13,Buy13}.
Light clusters, like deuterons or alpha particles, but also heavy
nuclei, which are embedded in the matter, will appear. In this case, 
clusters move freely and the matter can still be considered
homogeneous on length scales sufficiently larger than the cluster sizes.
At larger densities, just below nuclear saturation density, ``pasta
phases'' can arise, which, however, are not considered here. 
On the other hand, phase transitions with macroscopic regions of low and high
densities can occur, see Ref.\
\cite{Mue95,Gul13,Hem13,Zha13b}. Both these phenomena will be
considered in the present work.  They can have significant effects on the
density dependence of the symmetry energy.

In conventional model calculations the symmetry energy vanishes
linearly with decreasing density. Experimental results suggest considerably
higher values in the low-density limit \cite{Kow07,Nat10}
as compared to such a linear density dependence.
The modification of this behavior is described in a quantum statistical
approach as a result of cluster formation, which is, however, strongly
temperature dependent. With increasing density a good agreement with
other estimates is obtained \cite{Dan13}.

It is important to distinguish between finite nuclei, 
nuclear matter and stellar matter. The last two are infinite systems, 
where in nuclear matter the Coulomb interaction is turned off, 
while in stellar matter it is compensated 
by a lepton component requiring global charge neutrality. 
All these systems have substantially different phase
structures and chemical compositions.

The aim of the present work is the following.
After introducing the original definition of the symmetry energy in
nuclei, we consider those used in theoretical models for 
matter of finite density. These definitions are applied to calculations in
a theoretical model, which describes nuclear matter and stellar
matter in a consistent approach, and the resulting differences 
of different definitions are discussed. The effects of cluster correlations and of
the liquid-gas phase transition are
explored. Here the question of separating strong and electromagnetic
contributions to the energy will be seen to be important in order to allow for
meaningful comparisons and to be consistent with the separation of 
energy contributions for finite nuclei. 
This review is a continuation of our previous works on 
cluster correlations in matter \cite{Typ10,Vos12,Roe13}
with particular emphasis on the consequences 
for the symmetry energy in different systems. 

The content of this work is as follows: First, in section
\ref{sec:esym_th}, several quantities to characterize the variation of
the energy of a system as a function of the independent variables
are introduced. In particular, different theoretical definitions of the
symmetry energy in nuclei and infinite matter without cluster
formation will be considered.
In section \ref{sec:gRDF} a model for matter of finite density is presented that
is based on a generalized relativistic density functional (gRDF)
approach with nucleons, clusters and electrons (for stellar matter) as constituent
particles, where the medium properties of the clusters are calculated from a 
quantumstatistical model. The density functional approach
allows to study and to compare quantitatively the 
different definitions of the symmetry energy in
the considered systems under various thermodynamic conditions.
The conclusions are summarized in section \ref{sec:concl}.
Details about the liquid-gas phase transition construction, the medium
dependent mass shifts of the clusters and the effective degeneracy
factors of heavy nuclei can be found in the appendices
\ref{sec:lgpt}, \ref{sec:mshifts}, and \ref{sec:geff}, respectively.
Throughout this work we use
the traditional nuclear system of units where $\hbar = c = k_{B} = 1$.

\section{Definitions of the symmetry energy }
\label{sec:esym_th}

The definition of the symmetry energy depends on the type of 
system and the thermodynamic conditions. In the following, various
ways to introduce the symmetry energy and their relations will be
discussed.

\subsection{Nuclei}

For a nucleus with $N$ neutrons and $Z$
protons, the asymmetry $\delta$ is given by
\begin{equation}
 \delta = \frac{N-Z}{N+Z} \: ,
\end{equation}
which corresponds to the third component of the isospin $I_{z}$.
The binding energies $B(N,Z)$ of isobaric nuclei, i.e.\ nuclei with equal mass
number $A=N+Z$, show a characteristic
variation with the asymmetry $\delta$ that is almost symmetric with
respect to an exchange of neutrons and protons for light nuclei. 
The contribution of the electromagnetic interaction to the energy
leads to a violation of this symmetry, 
which is more severe for heavier nuclei with larger charge numbers. 
The isospin dependence is reflected in the semi-empirical or
Bethe-Weizs\"{a}cker mass formula \cite{Wei35,Bet36a,Lun03}
for nuclei. In this average description, the
binding energy of a nucleus is given by
\begin{eqnarray}
\label{eq:BW}
 B(N,Z) & = &  a_{V} A - a_{S} A^{2/3} 
 \\ \nonumber & & 
- \left( a_{V}^{(\rm sym)} A + a_{S}^{(\rm sym)} A^{2/3} \right)
 \delta^{2} 
 \\ \nonumber & & 
 - a_{C} \frac{Z(Z-1)}{A^{1/3}} + \dots 
\end{eqnarray}
with volume, surface, volume symmetry, surface symmetry and Coulomb
contributions that show a particular dependence on the mass number
$A$ and asymmetry $\delta$. The introduction of a surface symmetry term 
improves the description of the binding energies and allows to better separate the 
mass number dependence of the volume symmetry energy of infinite
matter. Various forms for the total symmetry energy are introduced in
the literature,
see e.g.\ Ref.\ \cite{Dan03}, which takes thermodynamic considerations into account.
Pairing and other features such as shell
effects are not considered here.
The parameters $a_{V}$, $a_{S}$, $a_{V}^{(\rm sym)}$, $a_{S}^{(\rm sym)}$, $a_{C}$ are found
by fitting nuclear masses
\begin{equation}
m_{N,Z} = N m_{n} + Z m_{p} - B(N,Z)
\end{equation}
across the whole chart of nuclei. Here $m_{n}$ and $m_{p}$ are the
neutron and proton rest masses, respectively. Typical values of the
coefficients are $a_{V} = 15.73$~MeV, $a_{S}=17.77$~MeV, $a_{V}^{(\rm
  sym)}=26.46$~MeV,
$a_{S}^{(\rm sym)}=-17.70$~MeV, and $a_{C}=0.709$~MeV
\cite{Lun03}.

\begin{figure}[t]
\begin{center}
\includegraphics[width=8.5cm]{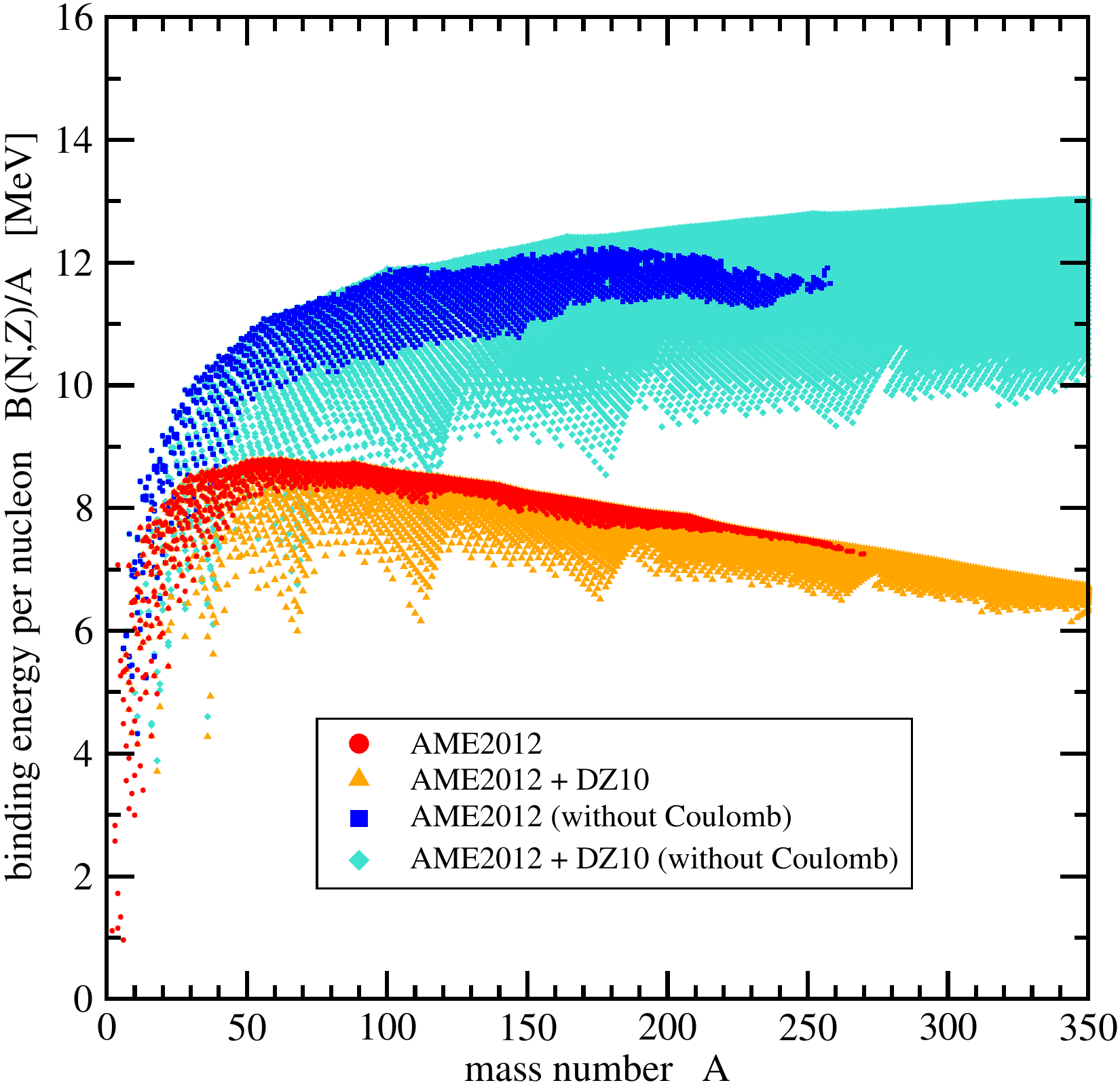}
\caption{\label{fig:bea_nuc_a}%
Binding energy per nucleon $B(N,Z)/A$ as a function of the mass number
$A$ for all nuclei
of the AME2012 atomic mass evaluation \cite{AME2012} and the
DZ10 mass model \cite{Duf95} with and without the Coulomb contribution 
$E_{\rm Coul}(N,Z)$. See text for details.}
\end{center}
\end{figure}

The binding energy per nucleon $B(N,Z)$
for four sets of nuclei is shown in figure \ref{fig:bea_nuc_a} 
as a function of the mass number
$A$. The first set (red circles) includes all nuclei with experimentally known
binding energies of the AME2012 atomic mass evaluation
\cite{AME2012}.
The second set (orange triangles) is an extension of the first set. It is obtained by
adding all nuclei with binding energy predictions of the
DZ10 mass formula \cite{Duf95}, which gives a rather good fit of the known masses.
All nuclei with $A\leq 350$ and positive neutron and proton separation
energies are included.
The usual pattern is observed with signs of shell closures, a maximum
in the iron region
and a smooth reduction of the binding energy per nucleon with
increasing mass number $A$ beyond the maximum. 
The third set (dark blue squares) covers the same nuclei as set 1
but the Coulomb energy
\begin{equation}
\label{eq:E_Coul}
 E_{\rm Coul} = a_{C} \frac{Z(Z-1)}{A^{1/3}}
\end{equation}
with $a_{C} = 3e^{2}/(5r_{0})$, is removed from the
binding energies of nuclei. Using the standard value $r_{0} = 1.25$~fm for the radius parameter,
we have $a_{C} = 0.6912$~MeV, which is slightly different from the values
derived in actual fits to binding energies.
The same transformation is applied to the
second set to obtain the fourth set (light blue diamonds).
The differences to the first two sets are
obvious. There is an continuous increase of the binding energy per
nucleon with $A$.
In sets 2 and 4, the variation of the $B(N,Z)/A$ for constant $A$
reflects the isospin dependence for isobaric nuclei.
A larger number of nuclei appears in the fourth set
since without the Coulomb contribution to the binding energy 
the neutron and proton driplines are shifted to much more exotic
nuclei, in particular for proton-rich nuclei.

\begin{figure}[t]
\begin{center}
\includegraphics[width=8.5cm]{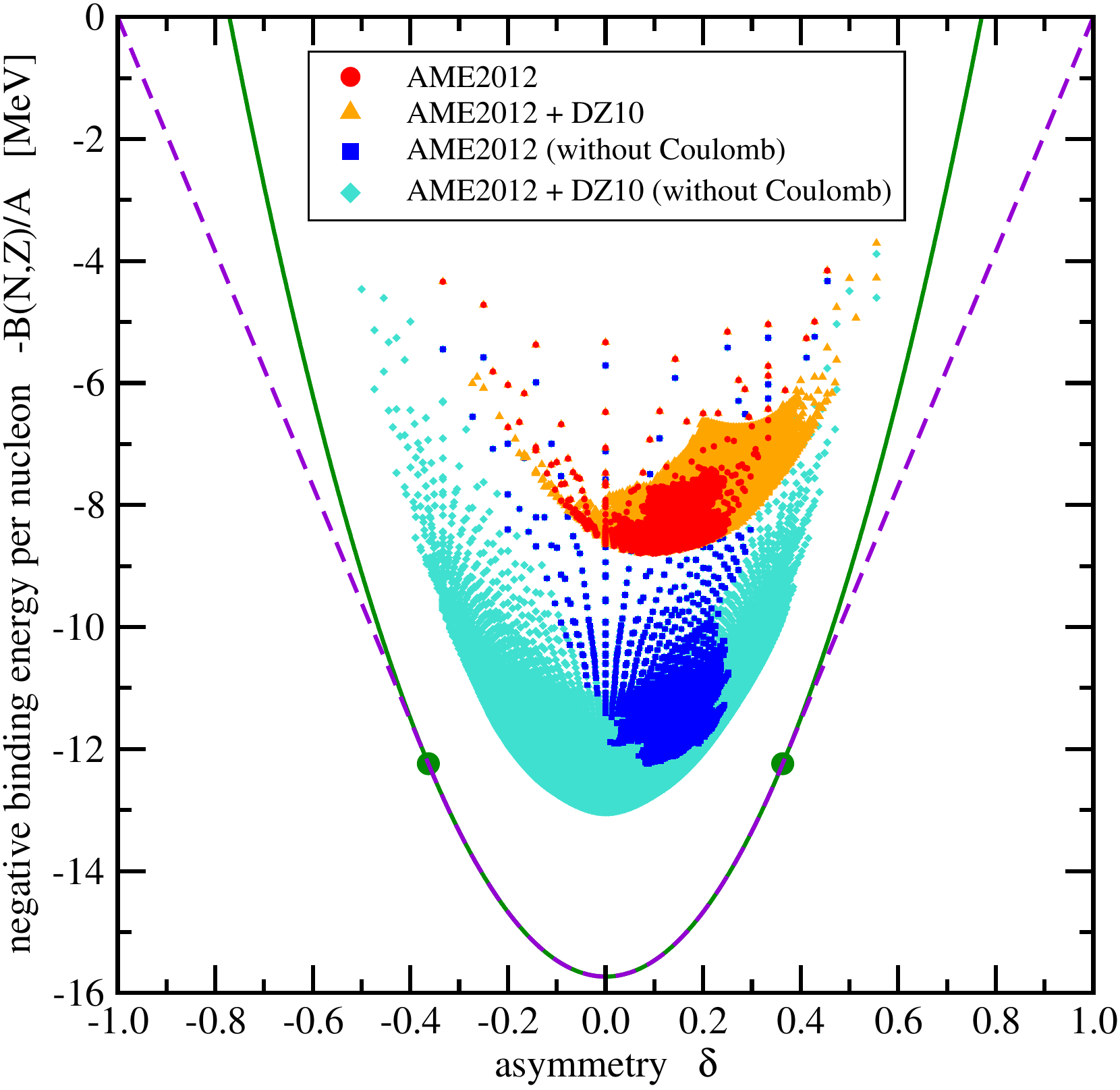}
\caption{\label{fig:bea_nuc_alpha}%
Negative binding energy per nucleon $-B(N,Z)/A$ as a function of the asymmetry $\delta$
for all nuclei with $A \leq 350$ of the atomic mass evaluation AME2012 and the
DZ10 mass formula \cite{Duf95} with and without the Coulomb contribution 
$E_{\rm Coul}(N,Z)$. The full green line is the infinite-$A$ limit
(\ref{eq:B_inf}) of the Bethe-Weizs\"{a}cker formula and the
green circles denote the neutron and proton drip points on this line.
The dashed violet line extends the full green curve in the range
$|\delta| \leq \delta_{\rm drip}$ by interpolating between the drip points and the case
$|\delta|=1$ with unbound neutrons and protons only. See text for details.}
\end{center}
\end{figure}

In figure \ref{fig:bea_nuc_alpha} the same sets of nuclei as in figure
\ref{fig:bea_nuc_a} are considered but now the negative binding energy
per nucleon is depicted as a function of the asymmetry parameter
$\delta$. The sets 1 and 2 that include the Coulomb contribution to the
binding energy show a minimum at $\delta \neq 0$ since the
isospin symmetry is clearly broken due to the Coulomb interaction.
When the Coulomb energy $E_{\rm Coul}$ is removed from $B(N,Z)$,
the distribution becomes more or less symmetric in $\delta$ with a
minimum at $\delta=0$. 
Constructing the lower bound of $-B(N,Z)$ for the nuclei of set 4, 
a piecewise linear function is obtained
that can be well approximated by a quadratic function of $\delta$  
close to the minimum.
Since the selection of nuclei in set 4 is
limited to those with mass number $A\leq 350$ the minimum curve does not 
represent the infinite nuclear matter result. The curve will
move downwards when more massive nuclei are included. 

\begin{figure}[t]
\begin{center}
\includegraphics[width=8.5cm]{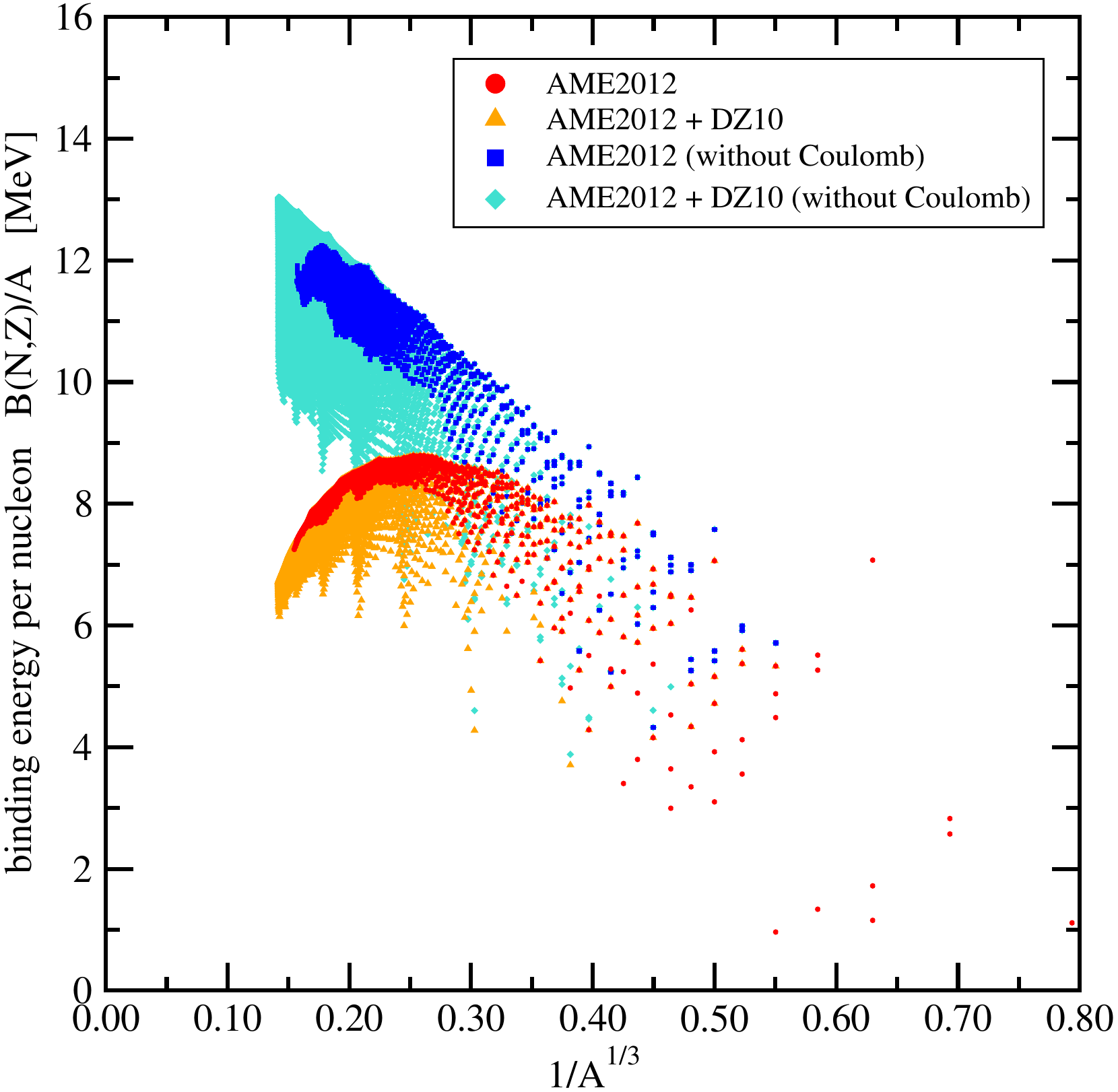}
\caption{\label{fig:bea_nuc_a_inv3}%
Binding energy per nucleon $B(N,Z)/A$ as a function of the 
inverse cubic root of the mass number, $1/A^{1/3}$, for all nuclei
with $A \leq 350$ of the AME2012 atomic mass evaluation \cite{AME2012} and the
DZ10 mass model \cite{Duf95} with and without the Coulomb contribution 
$E_{\rm Coul}(N,Z)$. For the latter one can extrapolate to the
infinite system and obtains the volume coefficient $a_{V}$ of the
symmetric matter binding energy functional. See text for details.}
\end{center}
\end{figure}

In the limit $A \to \infty$
and neglecting the Coulomb contribution to $B(N,Z)$,
only the two bulk contributions to the binding energy per nucleon
\begin{equation}
\label{eq:B_inf}
 B(N,Z)/A \to   B_{\infty}(\delta) = a_{V}  -  a_{V}^{(\rm sym)}   \delta^{2} \: ,
\end{equation}
remain. The corresponding curve (with $a_{V} = 15.73$~MeV and 
$a_{V}^{(\rm sym)}=26.46$~MeV) is shown in figure
\ref{fig:bea_nuc_alpha} by the full green line.
The minimum is given by $-a_{V}$ where the
coefficient $a_{V}$ is identified with the bulk binding energy
$B_{\rm  sat}$ of cold saturated infinite nuclear matter.
It can be obtained also by extrapolating the dependence of
$B(N,Z)/A$ on the inverse size of the system to $\lim_{A \to \infty}
1/A^{1/3} = 0$ for the case without the Coulomb contribution, see
figure \ref{fig:bea_nuc_a_inv3}.

Similarly, $a_{V}^{(\rm sym)}$ is the bulk nuclear symmetry energy at saturation 
and can be regarded as the symmetry energy of nuclear matter denoted below as $J$.
Note, that 
the parameter $a_{V}^{(\rm sym)}$ is found from fits to experimental data
to all nuclei, which follow only approximately equation (\ref{eq:BW}). It is neither
obtained from a second derivative nor from a finite difference of
energies as in the case of infinite matter (see below).

For asymmetries $\delta = \pm 1$, the quadratic form (\ref{eq:B_inf}) predicts
negative binding energies per nucleon since 
$a_{V} < a_{V}^{(\rm sym)}$, i.e.\ unbound systems. In fact, the curve should terminate 
at finite values of $\delta$ when the neutron and proton driplines are reached. In the limit
$A \to \infty$ this corresponds to the conditions 
\begin{equation}
 \left. \frac{d}{dN} \frac{B(N,Z)}{A} \right|_{\delta = \delta_{\rm drip}} = 0 
\end{equation}
and
\begin{equation}
 \left. \frac{d}{dZ} \frac{B(N,Z)}{A} \right|_{\delta = -\delta_{\rm drip}} = 0 \: 
\end{equation} 
that are obtained at the drip asymmetry $\delta_{\rm drip}$.
The value is found to be 
\begin{equation}
\label{eq:alpha_drip}
 \delta_{\rm drip} = 1 - \sqrt{1-\frac{a_{V}}{a_{V}^{(\rm sym)}}}
\end{equation}
with $\delta_{\rm drip}=0.3632$ for the above given parameters. 
These dripline points are denoted in figure \ref{fig:bea_nuc_alpha} by the green circles.

The most extreme values $\delta = \pm 1$ correspond to pure neutrons or protons with
zero binding energy. Connecting these points with the dripline points
on the full green line, the dashed violet line is obtained, which is a quadratic function for 
$|\delta| \leq \delta_{\rm drip}$ and a linear function for $|\delta| \geq \delta_{\rm drip}$.
This curve can be interpreted as the binding energy per nucleon of 
infinite nuclear matter in the zero-density limit at zero temperature. 
Since the energy per nucleon for $\delta = 0$ is non-vanishing, a finite symmetry energy
can be anticipated in this case, see subsection \ref{subsec:esym_gRDF}.
Following thermodynamical considerations, it is a convex function of $\delta$.
The transition asymmetry $\delta_{t}$ where the quadratic behavior changes to a linear one  
is easily found from the condition
\begin{equation}
 -\left.  \frac{dB_{\infty}}{d\delta} \right|_{\delta= \delta_{t}}
 = \frac{B_{\infty}(\delta_{t})}{1-\delta_{t}}
\end{equation}
to be identical with the drip asymmetry (\ref{eq:alpha_drip}).

Since a particular nucleus has a fixed number of neutrons and protons,
it is not reasonable to define the symmetry energy for a specific
nucleus itself. It is a property that characterizes the set of all nuclei.
However, if nuclei are studied in a hot and dense environment, their
energies are different as compared to those in vacuum and,
correspondingly, the coefficients in a generalized Bethe-Weizs\"{a}cker formula
can depend on the medium properties such as temperature and density
if such an description is applied.

\subsection{Infinite matter at finite density}

The concept of symmetry energy can also be applied to 
infinite matter at zero and finite temperature. 
However, there is a fundamental difference between nuclear
matter and stellar matter with considerably distinct phase diagrams. 
The former is a theoretical model for a
system of strongly interacting particles neglecting the
electromagnetic interaction because it gives a diverging
contribution in infinite systems. 
In contrast, for stellar matter that
occurs in compact stars or core-collapse supernovae,
both the strong and electromagnetic interactions are 
taken into account. The condition of global charge neutrality
requires to include charged leptons, in particular electrons. 
There are (at least) two independent conserved charges in infinite matter:
the total baryon number and
the total charge number. More exist, 
if strangeness and other lepton species are considered. Correspondingly, one can define
a total baryon number density $n_{B}$ and a charge number density $n_{Q}$.
For a system with $N_{n}$ neutrons and $N_{p}$ protons in a volume $V$, 
these densities are given by $n_{B} = (N_{n}+N_{p})/V$ and $n_{Q} =
N_{p}/V= (1-\delta) n_{B}/2$.
Then, the energy per particle $E$ can be expressed 
as a function of $n_{B}$ and $n_{Q}$
or, more conveniently, $n_{B}$ and the asymmetry $\delta$.
The functional dependence of the energy per particle on
$n_{B}$ and $\delta$ is also appropriate for
stellar matter since the number density of charged leptons $n_{L}$
(electrons and, at high densities, muons)
is determined via the charge neutrality condition.

\subsubsection{Nuclear matter}
\label{subsec:nuc_mat}

At low temperatures,
nuclear matter exhibits the phenomenon of saturation as a result of the
competition of long-range attraction and short-range repulsion between the
nucleons. Due to the isospin
symmetry of the strong interaction, the system assumes it lowest
energy for asymmetry $\delta=0$ and zero temperature at the saturation baryon density
$n_{\rm sat}$ if the neutron-proton mass difference is neglected. 
The energy per particle (without the rest mass contribution) 
in cold nuclear matter, i.e.\ at
zero temperature, can be expressed as
\begin{equation}
\label{eq:Ecoldnuclearmatter}
 E(n_{B},\delta) = E_{0}(n_{B}) + E_{\rm sym}(n_{B}) \: \delta^{2}
  + \mathcal{O}(\delta^{4})
\end{equation}
with the energy per particle of symmetric nuclear matter
$E_{0}(n_{B})= E(n_{B},0)$
and the density dependent symmetry energy
\begin{equation}
\label{eq:esym_def}
 E_{\rm sym}(n_{B}) = \frac{1}{2} \left.
 \frac{\partial^{2}}{\partial \delta^{2}} E(n_{B},\delta)
\right|_{\delta=0} \: ,
\end{equation}
sometimes denotes by $S(n_{B})$.
The quadratic behavior for the dependence on $\delta$ in 
eq.\ (\ref{eq:Ecoldnuclearmatter}) must be valid in the sense 
of an expansion in $\delta$ around zero due to the isospin symmetry 
of nuclear forces. It is often also well justified 
in homogeneous matter for larger ranges of $\delta$. Deviations
from the quadratic behavior are generally small except 
for baryon densities much different from the saturation density 
$n_{\rm sat}\approx 0.16$~fm.
 
It has been customary to expand the energy of symmetric nuclear matter
close to the saturation point as
\begin{equation}
 E_{0}(n_{B}) = - B_{\rm sat} + \frac{1}{2} K x^{2} + \frac{1}{6}
 K^{\prime} x^{3} + \dots
\end{equation}
for small $x=(n_{B}/n_{\rm sat}-1)/3$. There is no term linear in $x$ due
to the minimum condition at $x=0$. The coefficient $B_{\rm
  sat}=E_{0}(n_{\rm sat})$ is
identified with $a_{V}$ in the Bethe-Weizs\"{a}cker formula.
The coefficients
\begin{equation}
 K = 9 n_{B}^{2} \left. \frac{d^{2}}{dn_{B}^{2}} E_{0}(n_{B})
 \right|_{n_{B} = n_{\rm sat}}
\end{equation} 
and 
\begin{equation}
 K^{\prime} = 27  n_{B}^{3} \left. \frac{d^{3}}{dn_{B}^{3}} E_{0}(n_{B})
 \right|_{n_{B} = n_{\rm sat}}
\end{equation}
are the (in)compressibility of bulk nuclear matter
and the skewness coefficient, respectively.
In a similar way the symmetry energy admits an expansion 
\begin{equation}
\label{eq:esym_series}
 E_{\rm sym}(n_{B}) =  J + L x + \frac{1}{2}
   K_{\rm sym} x^{2} +   \dots 
\end{equation}
with the symmetry energy at saturation 
\begin{equation}
 J=E_{\rm sym}(n_{\rm  sat}) \equiv a_{V}^{(\rm sym)} \: ,
\end{equation}
the slope parameter
\begin{equation}
L = 3 n_{\rm sat} \left. \frac{d}{dn_{B}} E_{\rm sym}(n_{B}) 
 \right|_{n_{B}=n_{\rm sat}}
\end{equation}
and the symmetry curvature or symmetry incompressibility 
\begin{equation}
 K_{\rm sym} = 9 n_{B}^{2} \left. \frac{d^{2}}{dn_{B}^{2}} E_{\rm sym}(n_{B})
 \right|_{n_{B} = n_{\rm sat}} \: .
\end{equation}
It has been a major aim of recent efforts in nuclear physics to determine
values for these characteristic parameters of cold nuclear matter  
from various experiments, see, e.g., the contribution of
X. Vi\~{n}as et al.\ in this volume \cite{Vin13}.

Instead of the quantity $L$, the symmetry pressure $p_{0} = Ln_{\rm sat}/3$ is
introduced in some works.
In a different representation of 
the density dependence of the symmetry energy 
it is sometimes separated into a kinetic and a potential term as
\begin{equation}
\label{eq:esym_gamma}
 E_{\rm sym}(n_{B}) = 
 C_{\rm kin} \left( \frac{n_{B}}{n_{\rm sat}}\right)^{2/3}
 + C_{\rm pot}  \left( \frac{n_{B}}{n_{\rm sat}}\right)^{\gamma}
\end{equation}
with ($m_{\rm nuc} \approx m_{n} \approx m_{p} \approx 939$~MeV)
\begin{equation}
 C_{\rm kin} =  \frac{1}{6m_{\rm nuc}} 
 \left( \frac{3\pi^{2}}{2} n_{\rm sat} \right)^{2/3} \: . 
\end{equation}
The coefficients $C_{\rm pot}$ and $\gamma$ parametrize 
the density dependence of the symmetry energy in the region of
saturation density. At saturation we have
\begin{equation}
 C_{\rm pot} = J -C_{\rm kin} .
\end{equation}
The kinetic term originates from an expansion of the energy of a
free Fermi gas of neutrons and protons
\begin{eqnarray}
 \lefteqn{E_{\rm kin}(n_{B},\delta)}
 \\ \nonumber
  & = & \frac{3}{10m_{\rm nuc}} 
 \left( 3\pi^{2} n_{B} \right)^{2/3}
 \\ \nonumber & & \times
 \left[ \left(\frac{1+\delta}{2}\right)^{5/3} 
 + \left(\frac{1-\delta}{2} \right)^{5/3}
 \right]
 \\ \nonumber & = &  \frac{3}{10m_{\rm nuc}} 
 \left( \frac{3\pi^{2}}{2} n_{B} \right)^{2/3}
 \left[ 1 + \frac{5}{9}  \delta^{2} + \mathcal{O}(\delta^{4}) 
 \right] \: .
\end{eqnarray}
However, sometimes the effect of correlations has been introduced into
the kinetic energy term, see the contribution \cite{Car13} in this
volume and references therein,
e.g.\ by introducing an effective mass.
The parameter $\gamma$ in equation (\ref{eq:esym_gamma}) determines the slope 
\begin{equation}
 L = 2 C_{\rm kin} + 3\gamma C_{\rm pot}
 = (2-3\gamma) C_{\rm kin} + 3 \gamma J
\end{equation}
at the saturation density $n_{\rm sat}$. However, the slope $L$ is not
independent of the assumptions on $C_{\rm kin}$ and $J$. 
For $C_{\rm kin} = 12$~MeV, $J=32$~MeV,
and $\gamma = 1$ the value $L=84$~MeV is obtained for the slope parameter.

Nuclear matter can also be considered at finite temperatures $T$. From a
thermodynamic point of view, the appropriate quantity to be studied
is the free energy density $f(T,n_{B},n_{Q})$
or the free energy per particle $F(T,n_{B},\delta) 
 = f(T,n_{B},n_{Q})/n_{B}$ because
temperature and densities are the natural variables in this case.
The free energy per particle can be
expanded for small asymmetries as
\begin{equation}
\label{eq:Fnuclearmatter}
 F(T,n_{B},\delta) = F_{0}(T,n_{B}) + F_{\rm sym}(T,n_{B}) \: \delta^{2}
  + \mathcal{O}(\delta^{4})
\end{equation}
like in equation (\ref{eq:Ecoldnuclearmatter})
with the free energy per particle of symmetric nuclear matter
$F_{0}(T,n_{B})=F(T,n_{B},0)$ and the symmetry free energy 
\begin{equation}
\label{eq:fsym_def}
 F_{\rm sym}(T,n_{B}) = \frac{1}{2} \left.
 \frac{\partial^{2}}{\partial \delta^{2}} F(T,n_{B},\delta)
\right|_{\delta=0} \: .
\end{equation}
Both quantities are functions of temperature and baryon density.
For zero temperature the usual symmetry energy $E_{\rm sym}(n_{B})
= F_{\rm sym}(0,n_{B})$ is recovered.

Instead of the free energy per particle (\ref{eq:Fnuclearmatter}), one
can consider the internal energy per particle $U$. From the standard relations 
of thermodynamics between free and internal energies 
with the entropy per particle $S$ we obtain the symmetry internal energy
per particle
\begin{equation} 
\label{eq:usym_def}
 U_{\rm sym}(T,n_{B}) = F_{\rm sym}(T,n_{B}) 
 + T S_{\rm sym}(T,n_{B})
\end{equation}
with the symmetry entropy per particle
\begin{eqnarray}
 S_{\rm sym}(T,n_{B}) & = & - \frac{1}{2} \left.
 \frac{\partial^{2}}{\partial \delta^{2}} \frac{\partial}{\partial T}
 F(T,n_{B},\delta)
\right|_{\delta=0} 
 \\ \nonumber  & = &
 - 
 \frac{\partial}{\partial T} F_{\rm sym}(T,n_{B})
 \: .
\end{eqnarray}
Hence, there is a difference between symmetry
free energies and symmetry internal energies, 
which can be substantial for large temperatures.
However, another definition of the symmetry internal energy can be
introduced because
$T$ is not the natural variable of the thermodynamic potential $U$.
In fact, the natural
variables are the entropy per particle $S$ and the baryon number and charge number
densities. Then, the expansion for small symmetries reads
\begin{equation}
\label{eq:Unuclearmatter}
 U(S,n_{B},\delta) = U_{0}(S,n_{B}) + U_{\rm sym}(S,n_{B}) \: \delta^{2}
  + \mathcal{O}(\delta^{4})
\end{equation}
with the symmetry internal energy
\begin{equation}
\label{eq:Usym_deriv}
 U_{\rm sym}(S,n_{B}) = \frac{1}{2} \left.
 \frac{\partial^{2}}{\partial \delta^{2}} U(S,n_{B},\delta)
\right|_{\delta=0} \: .
\end{equation}
Employing standard thermodynamic identities, the relations
\begin{equation}
 \left. \frac{\partial}{\partial \delta} F(T,n_{B},\delta)
   \right|_{T,n_{B}} =
 \left. \frac{\partial}{\partial \delta} U(S,n_{B},\delta) \right|_{S,n_{B}}
\end{equation}
and 
\begin{eqnarray}
 F_{\rm sym}(T,n_{B})
& = &  U_{\rm sym}(S,n_{B})
 \\ \nonumber & & 
 +  \frac{1}{2} \left. \frac{\partial T}{\partial \delta}  \right|_{S,n_{B},\delta=0}
 \left. \frac{\partial S}{\partial \delta} \right|_{T,n_{B},\delta=0}
\end{eqnarray}
are found.
Obviously, the isospin dependence of the energy in systems with equal
entropy per particle and baryon density has to be distinguished from that
of systems with equal temperature and baryon density.  
In the following, only symmetry energies in systems of constant
temperature will be considered.

The definitions (\ref{eq:esym_def}), (\ref{eq:fsym_def}), 
and (\ref{eq:Usym_deriv})
using second derivatives are motivated by the
expansion of the energy per particle in nuclear matter near saturation
density in apower series in $\delta$.
Assuming a quadratic dependence on the asymmetry $\delta$
for the whole range of $\delta$
these definitions can be replaced by finite difference formulas, e.g.
\begin{eqnarray}
\label{eq:findiff}
 \lefteqn{E_{\rm sym}(n_{B})}
 \\ \nonumber & = & \frac{1}{2} \left[ E(n_{B},+1)- 2E(n_{B},0) + E(n_{B},-1)\right]
\end{eqnarray}
and similar for the free energy per particle $F$ and the internal
energy per particle $U$ with identical results for the two
definitions. 
In this case, the energy of symmetric matter
($\delta=0$) is compared to pure neutron ($\delta=1$) and pure proton ($\delta =
-1$) matter. In many cases, however, there is a difference of these
definitions using second derivatives or finite differences due to
deviations from the $\delta^{2}$ dependence of the energies 
for finite values of $\delta$ as will be
shown below and was discussed already in Ref.\ \cite{Typ10}. 
The derivative is not always well defined and can diverge if, e.g.,
light clusters at very low temperatures are considered in the model,
see Ref.\ \cite{Typ10}.
The finite-difference formula (\ref{eq:findiff}) is always
applicable. We will show that it
represents the effect of the isospin dependence adequately.

For low temperatures $T$, the pressure 
\begin{equation}
 p(T,n_{B},\delta) = n_{B}^{2} \left.
 \frac{\partial F}{\partial n_{B}} \right|_{T,\delta}
\end{equation}
can become negative for densities $n_{B}$ lower than the saturation
density $n_{\rm sat}$. This behavior indicates that nuclear matter
becomes unstable against density fluctuations in this region.
As a consequence, the system will no longer remain spatially homogeneous.
Finite-size clusters will form separating regions of high and low
densities. Since an increase of the cluster size will lead to a
larger binding energy per particle in general, cf.\ the Bethe-Weizs\"{a}cker
formula
(\ref{eq:BW}) without the Coulomb contribution, 
the whole system will separate into two
macroscopic phases in the infinite volume limit
and the well-known liquid-gas phase transition will
emerge. The correct state in thermodynamic equilibrium can be found,
e.g., by minimizing the free energy density $f(T,n_{B},n_{Q})$ globally.
Since there are two conserved charges, corresponding to the
independent particle densities
$n_{B}$ and $n_{Q}$, in the system, a ``noncongruent'' phase
transition is the result
\cite{Mue95,Gul13,Hem13,Zha13b}.
The thermodynamic quantities in the region of coexisting phases
can be constructed using the well-known Gibbs conditions, i.e.\
the equality of all intensive thermodynamic variables in all
phases. See appendix \ref{sec:lgpt} for details.

\begin{figure}[t]
\begin{center}
\includegraphics[width=8.5cm]{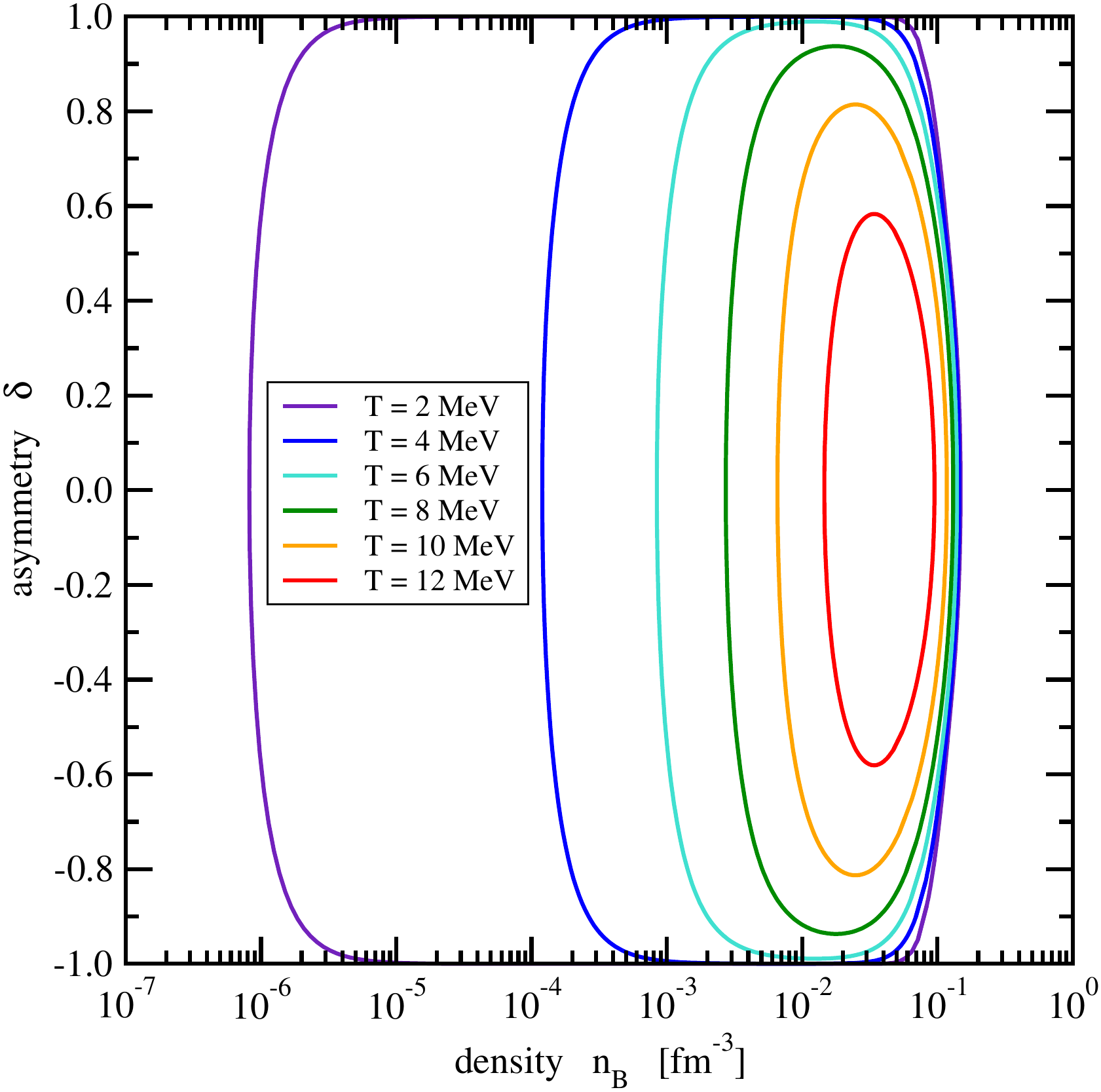}
\caption{\label{fig:binodals_log}%
Binodals of nuclear matter in the gRDF model for various temperatures $T$ in the 
asymmetry-density plane. Cluster formation is not taken into account.}
\end{center}
\end{figure}

The boundaries of the phase coexistence region or ``binodals''
are depicted for various temperatures in figure \ref{fig:binodals_log}
for the gRDF model of section \ref{sec:gRDF} without clusters. 
With decreasing temperature,
the enclosed coexistence region grows and covers larger sections of the
asymmetry-density plane; with larger asymmetry the region of
coexistence shrinks. Above the critical temperature 
$T_{\rm  crit}=13.724$~MeV of the gRDF model for nuclear matter, 
there is no separation of phases. 
The critical baryon density for the gRDF model without clusters is 
$n_{B}^{(\rm crit)}=0.04515$fm$^{-3}$.
Due to a finite symmetry energy, 
the isospin asymmetries in the two coexisting phases will be different
for a system with asymmetry $\delta\neq 0$. The high-density
phase (liquid) is found to be more isospin symmetric 
and the low-density phase (gas) more isospin asymmetric. 
This behavior has sometimes been called isospin destillation or
fractionation, see, e.g., Ref.\ \cite{Col08}.
The existence of the liquid-gas phase transition in nuclear
matter affects the density dependence of the symmetry energy. This
will be studied in section \ref{sec:gRDF}.

\subsubsection{Stellar matter}

Stellar matter in compact stars or core-collapse supernovae
represents a system containing strongly and
electromagnetically interacting
particles. It is considerably different from nuclear matter in
respect to thermodynamic properties. 
The charge of particles cannot be neglected and the condition of total
charge neutrality requires to include electrons as constituent
particles. They form a homogeneous distribution of degenerate
fermions at sufficiently high densities. 
Since the electronic background is not completely incompressible,
the interaction between electrons and charged baryonic particles
can induce electron-cluster correlations with a local increase of the
electron density near clusters.
This effect can be treated approximately in calculations by employing the Wigner-Seitz
approximation.

Correlations due to the strong and electromagnetic interactions affect the phase
structure of the system in a different way as compared to nuclear
matter. The competition between the attractive nuclear interaction and the
repulsive Coulomb interaction favors the formation of finite-size
structures \cite{Gul13}, e.g.\ clusters. 
The Coulomb contribution to the cluster energy is
screened by the electron background leading to an increased binding. 
Considering the full table of nuclei, this causes
a shift of the position of the most-bound nucleus to larger mass numbers.
Furthermore, the properties of nuclei in the medium are
modified as compared to their vacuum values as a consequence of the
Pauli exclusion principle and the impact of the nuclear interaction. 
Note that the results for nuclear matter in the previous subsection
were obtained without cluster formation.

The definitions of the symmetry energy that were introduced for
infinite nuclear matter can be transfered directly to the stellar
matter case. However, since the contribution of electrons and the
electromagnetic interaction are included,
the energy of the matter does not show a simple isospin symmetry any more.
At very low temperatures and not too high densities,
stellar matter undergoes a phase transition to a Wigner crystal
indicating the importance of long-range Coulomb correlations.
At higher densities, so-called ``pasta'' phases appear.
Due to all these features, the extraction of a symmetry energy
in stellar matter will yield results that can be different to nuclear
matter. Appropriate modifications have to be applied in order to give comparable
quantities, in particular a correction for the Coulomb contribution.

In core-collapse supernovae, the properties of stellar matter are
probed in a large range of asymmetries $\delta$. In cold compact stars,
however, stellar matter will be in $\beta$ equilibrium and the
asymmetry is fully determined due to the simultaneously
required condition of charge neutrality. Assuming an ideal mixture of
(interacting) nucleons and relativistic electrons, the total energy density at
zero temperature is given by
\begin{eqnarray}
 \lefteqn{\varepsilon_{\rm tot}(n_{n},n_{p},n_{e})}
 \\ \nonumber & = & m_{n} n_{n} + m_{p} n_{p}
 + E(n_{B},\delta) n_{B} 
 + \frac{3}{4}E_{e} n_{e} + \frac{1}{4} m_{e} n_{e}^{(s)} 
\end{eqnarray}
with
\begin{equation}
 E_{e} = \sqrt{k_{e}^{2}+m_{e}^{2}} \: ,
\end{equation}
the electron Fermi momentum 
\begin{equation}
 k_{e} = \left(3\pi^{2}n_{e}\right)^{1/3} \: ,
\end{equation}
which depends on the electron density $n_{e}$,
and the scalar electron density
\begin{equation}
 n_{e}^{(s)} = \frac{m_{e}}{2\pi^{2}} \left[ k_{e} E_{e}
 - m_{e}^{2} \ln \frac{k_{e}+E_{e}}{m_{e}}\right] \: .
\end{equation}
The condition of $\beta$ equilibrium can be expressed as
\begin{equation}
 \mu_{n} = \mu_{p} + \mu_{e}
\end{equation}
with the chemical potentials 
\begin{equation}
 \mu_{i} = \left. \frac{\partial \varepsilon_{\rm tot}}{\partial
     n_{i}} \right|_{n_{j \neq i}}
\end{equation}
of the particles $i=n,p,e$.
Considering charge neutrality, i.e.\ $n_{e} = n_{p}= (1-\delta)n_{B}/2$ and
assuming in eq.\ (\ref{eq:Ecoldnuclearmatter}) a purely quadratic dependence
of the nuclear matter energy  $E(n_{B},\delta)$ 
on $\delta$ (without the electronic
contribution),
the asymmetry $\delta_{\beta}(n_{B})$ at $\beta$ equilibrium is determined by the
condition (c.f.\ Ref.\ \cite{Kla06})
\begin{equation}
 4\delta_{\beta}E_{\rm sym}(n_{B}) +m_{n} - m_{p} =
 \mu_{e}(n_{B},\delta_{\beta}) 
\end{equation}
with 
\begin{equation}
 \mu_{e}(n_{B},\delta_{\beta}) = E_{e}
\end{equation}
and thus determined by the density dependence of the symmetry energy.

\section{Generalized relativistic density functional for dense matter}
\label{sec:gRDF}

In order to illustrate the effects of the liquid-gas phase transition
in nuclear matter without clustering and of cluster formation in stellar matter on
the symmetry energy, a theoretical model is required that is able to 
describe these features. We adopt, extend and modify the approach of
Refs.\ \cite{Typ10,Vos12} that is based on a relativistic mean-field model for
nuclear matter and nuclei with
density dependent meson-nucleon couplings. Besides nucleons, light and heavy
clusters are included as degrees freedom with medium dependent
properties.

\subsection{Thermodynamic quantities}
\label{subsec:gRDF_thermo}

Using a grand canonical description, all relevant thermodynamic
quantities can be derived from the grand canonical potential density
$\omega_{gc}(T,\mu_{i})$ 
that is a function of the temperature $T$ and
chemical potentials $\mu_{i}$ of all constituents. 
In the case of
nuclear matter, only neutrons ($i=n$) and protons ($i=p$) are considered. For stellar
matter, also nuclei ($i=(N,Z)$) and electrons ($i=e$)
are included as degrees of freedom.
In order to reproduce the model independent virial equation of state
at low baryon densities, see. e.g. \cite{Hor06a,Hor06b,OCo07}, 
also two-nucleon continuum correlations
($i=(nn)_{T=1}$, $(np)_{T=1}$, $(np)_{T=0}$, $(pp)_{T=1}$) 
in the appropriate isospin channels are 
introduced and represented by effective medium dependent cluster resonance
states. The effects of the strong interaction are modeled by the
exchange of effective mesons ($m=\sigma,\omega,\rho$) that couple
minimally to the free nucleons and nucleons in clusters. 

A baryon number $B_{i}$, a
charge number $Q_{i}$ and an (electronic) lepton number $L_{i}$ are
associated to each particle $i$. For clusters, we have $B_{i} = N_{i}+Z_{i} = A_{i}$,
$Q_{i} = Z_{i}$, and $L_{i}=0$ whereas for electrons $B_{e}=0$, $Q_{e} =
-1$, and $L_{e} = 1$.

Every particle with rest mass $m_{i}$ in the vacuum is considered
as a quasiparticle with energy
\begin{equation}
\label{eq:E_qp}
 E_{i}(k) = \sqrt{k^{2}+\left(m_{i}-S_{i} \right)^{2}}+V_{i}
\end{equation}
that depends on the particle momentum $k$, the scalar potential
$S_{i}$, and the vector potential $V_{i}$. These potentials contain
effects of the strong and electromagnetic interaction and of the medium modification of
particle properties. 

Experimental rest masses are used for neutrons, protons and electrons.
The rest masses of clusters are given by
\begin{equation}
 m_{i} = N_{i} m_{n} + Z_{i} m_{p} - B(N_{i},Z_{i})
\end{equation}
with vacuum binding energy $B(N_{i},Z_{i})$. For nuclei they are taken from
the atomic mass evaluation AME2012 \cite{AME2012} if they are experimentally known.
For other nuclei the values of the DZ10 \cite{Duf95}
mass formula are assumed. For
the two-nucleon resonance states we have $B(N_{i},Z_{i})=0$.

The scalar potential
\begin{equation}
 S_{i} =  \Gamma_{i\sigma} A_{\sigma} - \Delta m_{i}
\end{equation}
 of a particle $i$
and the vector potential
\begin{equation}
\label{eq:Vi}
 V_{i} =  \Gamma_{i\omega} A_{\omega} + \Gamma_{i\rho} A_{\rho}
 + V_{i}^{(r)}
\end{equation}
appearing in equation (\ref{eq:E_qp})
receive contributions from the meson fields with strengths $A_{m}$
($m=\sigma,\omega,\rho$)
and couplings $\Gamma_{im}$ which are a product
\begin{equation}
 \Gamma_{im} = g_{im} \Gamma_{m}(n_{B}) 
\end{equation}
of constant scaling factors $g_{im}$ and functions $\Gamma_{m}(n_{B})$
of the total baryon number density 
\begin{equation}
\label{eq:n_B}
 n_{B} = \sum_{i} B_{i} n_{i} \: .
\end{equation}
We use $g_{i\sigma} = g_{i\omega} = B_{i}$ and $g_{i\rho} =
B_{i}-2Q_{i}$ for nucleons and clusters, i.e.\ nucleons bound in
nuclei couple with the same strength to mesons as free nucleons.
Obviously, $g_{e\sigma} = g_{e\omega} = g_{e\rho} = 0$ for electrons.
The functional dependence of the couplings $\Gamma_{m}$ 
has the form as given in Ref.\ \cite{Typ10} 
with the well-calibrated DD2 parametrization that
was obtained from a fit to finite nuclei.
The mass shifts $\Delta m_{i}$ contributing to the mean-field
contribution in the scalar potential $S_{i}$ are
medium-dependent properties determined by the Pauli blocking. 
They are specified in detail in appendix
\ref{sec:mshifts}. For nucleons and electrons $\Delta_{i}=0$.

The contributions to the grand canonical potential density 
\begin{eqnarray}
\label{eq:omega_gc}
 \omega_{gc}(T,\mu_{n}, \mu_{p}, \dots) & = &
 \sum_{i} \omega_{i}^{(\rm qp)} + \omega^{(\rm int)} 
\end{eqnarray}
are those of independent quasiparticles
\begin{eqnarray}
\label{eq:omega_qp}
 \lefteqn{\omega_{i}^{(\rm qp)}}
 \\ \nonumber & = & - g_{i} \frac{T}{\sigma_{i}} \int \frac{d^{3}k}{(2\pi)^{3}}
 \ln \left\{ 1 + \sigma_{i} \exp 
\left[ - \frac{E_{i}(k)-\mu_{i}}{T}\right] \right\}
\end{eqnarray}
(with $\sigma_{i} = +1$ for fermions and $\sigma_{i} = -1$ for bosons)
and that of the interaction
\begin{eqnarray}
\label{eq:omega_int}
 \omega^{(\rm int)} & = & 
 \frac{1}{2} m_{\sigma}^{2} A_{\sigma}^{2}
 - \frac{1}{2} m_{\omega}^{2} A_{\omega}^{2}
 - \frac{1}{2} m_{\rho}^{2} A_{\rho}^{2}
 \\ \nonumber & &
  - \sum_{i} V_{i}^{(r)} n_{i}
\end{eqnarray}
with rearrangement potentials $V_{i}^{(r)}$ which also
appear in the vector potentials (\ref{eq:Vi}). They are required in
order to have a thermodynamic consistent theory (see below).

The quantity $g_{i}$ in equation (\ref{eq:omega_qp}) denotes the
degeneracy factor of a particle $i$. For nucleons and light nuclei we have $g_{p} =
g_{n} = 2$ and $g_{(1,1)}=3$ (${}^{2}$H), $g_{(2,1)} = g_{(1,2)} = 2$
(${}^{3}$H, ${}^{3}$He), $g_{(2,2)}=1$ (${}^{4}$He). For heavier
nuclei, $g_{(N,Z)}$ depends on the temperature $T$ due to the
excitation of states in a warm medium, see appendix \ref{sec:geff}. 
The effective temperature dependent degeneracy factors for the
two-nucleon resonance states are determined from the consistency
relations as discussed in Ref.\ \cite{Vos12}.

The single-quasiparticle number densities $n_{i}$ in equations
(\ref{eq:n_B}) and (\ref{eq:omega_int}) are given by
\begin{equation}
\label{eq:n_i}
 n_{i}   = 
  g_{i} \int \frac{d^{3}k}{(2\pi)^{3}} \:  f_{i}(E_{i},\mu_{i},T)
\end{equation}
with the convential distribution functions
\begin{equation}
\label{eq:distr}
 f_{i}(E_{i},\mu_{i},T) = \left[ \exp\left( \frac{E_{i}-\mu_{i}}{T}\right)
 + \sigma_{i}\right]^{-1} 
\end{equation}
for fermions  and bosons. 
Note that the grand canonical potential density (\ref{eq:omega_gc})
is a functional of temperature and chemical potentials even though the
densities $n_{i}$ appear explicitly in the definition of the
individual contributions to $\omega_{gc}$.
The form (\ref{eq:n_i}) is consistent with the thermodynamic
definition
\begin{equation}
 n_{i} = - \left. \frac{\partial \omega_{gc}}{\partial \mu_{i}}
 \right|_{T,\mu_{j}\neq \mu_{i}} 
\end{equation}
if the rearrangement potentials are defined correctly.
For nuclei with mass number $A>4$ it is sufficient to use
Maxwell-Boltzmann statistics, corresponding to the limit $\sigma_{i}
\to 0$ in equations (\ref{eq:omega_qp}) and (\ref{eq:distr}).
Furthermore we use for these nuclei the nonrelativistic approximation
(including rest mass)
\begin{equation}
 E_{i}(k) = \frac{k^{2}}{2(m_{i}-S_{i})}+m_{i}-S_{i} + V_{i}
\end{equation}
of the quasiparticle energies (\ref{eq:E_qp}). Then we find
\begin{equation}
  \omega_{i}^{(\rm qp)} = -Tn_{i} = -\frac{g_{i}T}{\lambda_{i}^{3}}
  \exp\left( \frac{\mu_{i}-V_{i}+S_{i}}{T}\right)
\end{equation}
with the thermal wavelengths
$\lambda_{i} = \sqrt{2\pi/[(m_{i}-S_{i})T]}$.

The strengths $A_{m}$ of the meson fields appear as auxiliary
quantities in the grand canonical potential density
$\omega_{gc}$. They are obtained from the (trivial) fields equations
\begin{equation}
\label{eq:meson}
 m_{m}^{2} A_{m} = \Gamma_{m}^{2} n_{m}
\end{equation}
that are found with the help of the Euler-Lagrange equations.
The source densities $n_{m}$ in equation (\ref{eq:meson}) are given by
\begin{eqnarray}
 n_{\omega} & = & \sum_{i} g_{i\omega} n_{i}
 \\
 n_{\rho} & = & \sum_{i} g_{i\rho} n_{i}
 \\
 n_{\sigma} & = & \sum_{i} g_{i\sigma} n_{i}^{(s)}
\end{eqnarray}
with the scalar quasiparticle number densities
\begin{eqnarray}
 \lefteqn{n_{i}^{(s)}} \\ \nonumber  & = & 
 g_{i} \int \frac{d^{3}k}{(2\pi)^{3}} \:  f_{i}(E_{i},\mu_{i},T)
 \frac{m_{i}-S_{i}}{\sqrt{k^{2}+(m_{i}-S_{i})^{2}}} \: .
\end{eqnarray}
This integral reduces to
\begin{equation}
 n_{i}^{(s)} = n_{i} \left(1-\frac{3}{2} \frac{T}{m_{i}-S_{i}} \right)
\end{equation}
for heavy nuclei with $A>4$ in the above-mentioned approximation.

The rearrangement potentials
\begin{equation}
 V_{i}^{(r)} = B_{i} U^{(\rm meson)} + U_{i}^{\rm (mass)} 
\end{equation}
include the standard meson contribution
\begin{equation}
 U^{(\rm meson)} 
 = \Gamma_{\omega}^{\prime} A_{\omega} n_{\omega}
 + \Gamma_{\rho}^{\prime} A_{\rho} n_{\rho}
 - \Gamma_{\sigma}^{\prime} A_{\sigma} n_{\sigma} 
\end{equation}
with derivatives
$\Gamma_{m}^{\prime} = d\Gamma_{m}/d\varrho_{V}$ of the couplings
and a term
\begin{equation}
 U_{i}^{\rm (mass)} = \sum_{j}  \frac{\partial \Delta
   m_{j}}{\partial n_{i}} n_{j}^{(s)}
\end{equation}
related to the medium dependent mass shifts of the quasiparticles.

The entropy density is obtained from the thermodynamic definition
\begin{eqnarray}
 s & = & - \left.\frac{\partial}{\partial T}
   \omega_{gc}\right|_{\mu_{i}}
 \\ \nonumber  & = & 
 - \sum_{i} g_{i} \int \frac{d^{3}k}{(2\pi)^{3}}
 \left[ f_{i} \ln \left(f_{i}\right)
 \right. \\ \nonumber & & \left.
+ \sigma_{i} \left(1 -  \sigma_{i} f_{i} \right) \ln \left( 1 - \sigma_{i}
     f_{i}\right)
 \right]
\\ \nonumber & & 
 -  \sum_{i} \left[ 
 \frac{d \ln (g_{i})}{d T} 
 \omega_{i}^{\rm (qp)}
 +  \frac{\partial \Delta m_{i}}{\partial T}   n_{i}^{(s)}
 \right]
\end{eqnarray}
with two non-standard terms in addition to the 
conventional contribution. 
They are caused by the temperature dependence of the 
degeneracy factors and of the mass shifts. 
Further thermodynamic quantities such as
the free energy density
\begin{equation}
\label{eq:f_dens}
 f = \omega_{gc} + \sum_{i} \mu_{i} n_{i}
\end{equation}
and the internal energy density
\begin{equation}
\label{eq:u_dens}
 u = f + T s
\end{equation}
are immediately obtained from the grand canonical potential density
$\omega_{gc}$ which is just the negative pressure $p$.

We study dense matter in chemical equilibrium and assume
that all reactions that change the chemical composition of the system, 
except those mediated by the weak interaction,
are equilibrated. As a result, the chemical potentials $\mu_{i}$ of
all particles are not independent. Because there are three independent
conserved charges (baryon number, charge number, lepton number) the
corresponding three
chemical potentials $\mu_{B}$, $\mu_{Q}$, $\mu_{L}$ are sufficient to
specify the chemical potentials
\begin{equation}
\label{eq:mu_part}
 \mu_{i} = B_{i} \mu_{B} + Q_{i} \mu_{Q} + L_{i} \mu_{L}
\end{equation}
for all constituents. In nuclear matter, leptons are not considered
and the leptonic contribution in (\ref{eq:mu_part}) can be ignored. 
In this case, the total charge number density
\begin{equation}
 n_{Q} = \sum_{i} Q_{i} n_{i} \geq 0
\end{equation}
is related to the asymmetry by 
\begin{equation}
 n_{Q} = \frac{1-\delta}{2}n_{B} \: .
\end{equation}
In stellar matter, there is the additional condition of charge neutrality
$n_{Q} = 0$
that determines the (electronic) lepton chemical potential $\mu_{L}$
for given asymmetry 
\begin{equation}
 \delta = 1- 2\frac{n_{L}}{n_{B}} 
\end{equation}
of the matter with 
\begin{equation}
 n_{L} = \sum_{i} L_{i} n_{i} = n_{e} \geq 0
\end{equation} 
when only electrons are considered.
Thus, there are only two
independent chemical potentials $\mu_{B}$ and $\mu_{Q}$.

\subsection{Symmetry energy in the gRDF approach}
\label{subsec:esym_gRDF}

The density dependence of the symmetry energy below nuclear saturation
density will be affected by two
reasons: the definition of the symmetry energy and the occurrence of 
spatial inhomogeneities be it a phase transition or the appearance of
finite size clusters. In this subsection, the effects will be
presented as they appear by applying the gRDF model. 
For the DD2 parametrization, used in the present
calculations, the saturation density of symmetric nuclear matter at
zero temperature is $n_{\rm sat} \approx 0.149$~fm${}^{-3}$
\cite{Typ10}. Above this density, no effects from the
liquid-gas phase transition or cluster formation occur and the
usual results for the density dependence of the symmetry energy
are recovered. Hence, we
limit the range in the figures to sub-saturation densities.

\begin{figure}[t]
\begin{center}
\includegraphics[width=8.5cm]{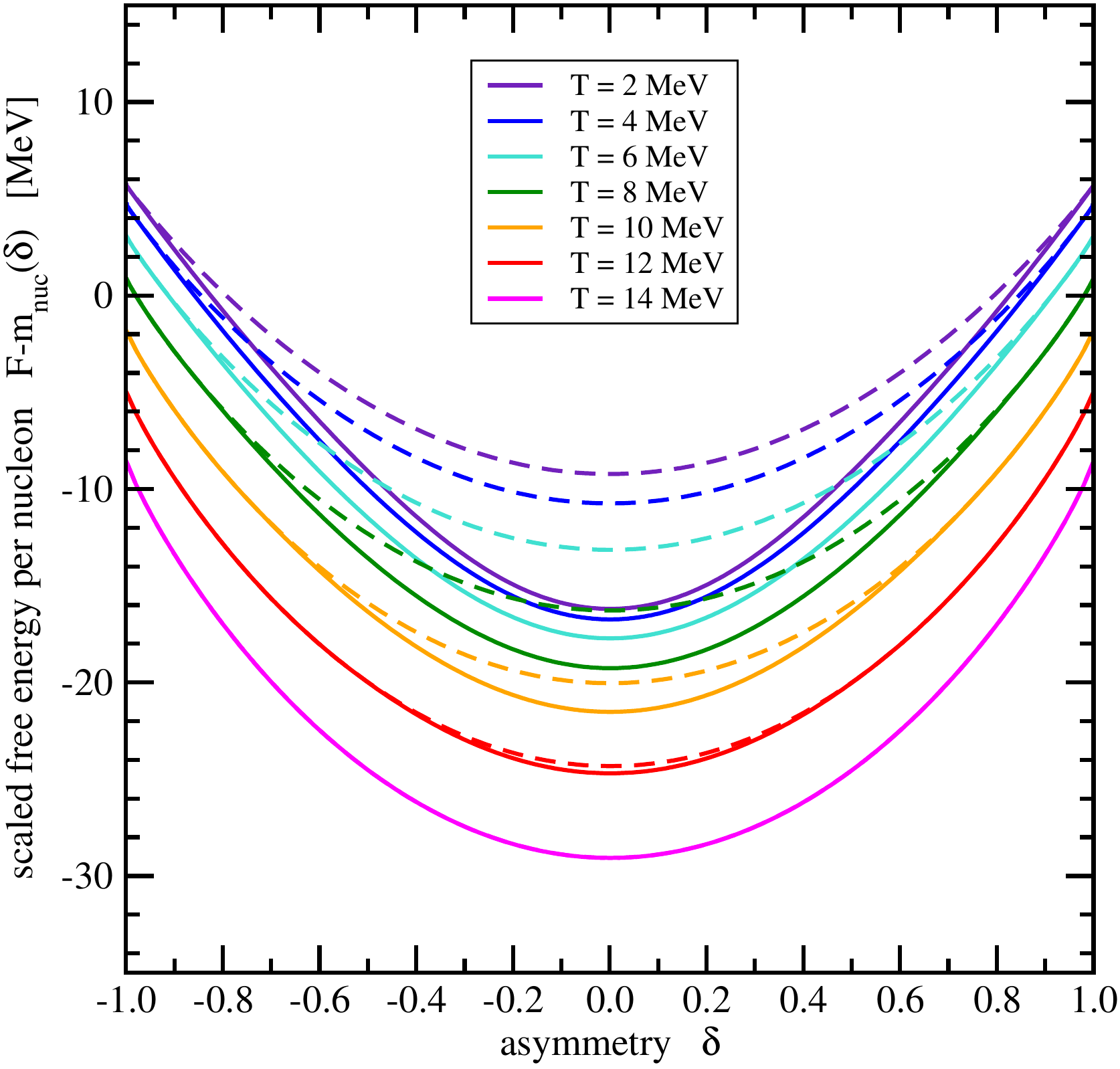}
\caption{\label{fig:F_alpha}%
Free energy per nucleon $F$ corrected for the trivial mass
contribution $m_{\rm nuc}(\delta)$, Eq.\ (\ref{eq:m_nuc}), 
as a function of the asymmetry
$\delta$ in nuclear matter at the critical baryon density $n_{\rm
  crit}$ of the DD2 parametrization for different temparatures $T$
without (dashed lines) and with (full lines) liquid-gas phase
transition without cluster formation.}
\end{center}
\end{figure}

\begin{figure}[t]
\begin{center}
\includegraphics[width=8.5cm]{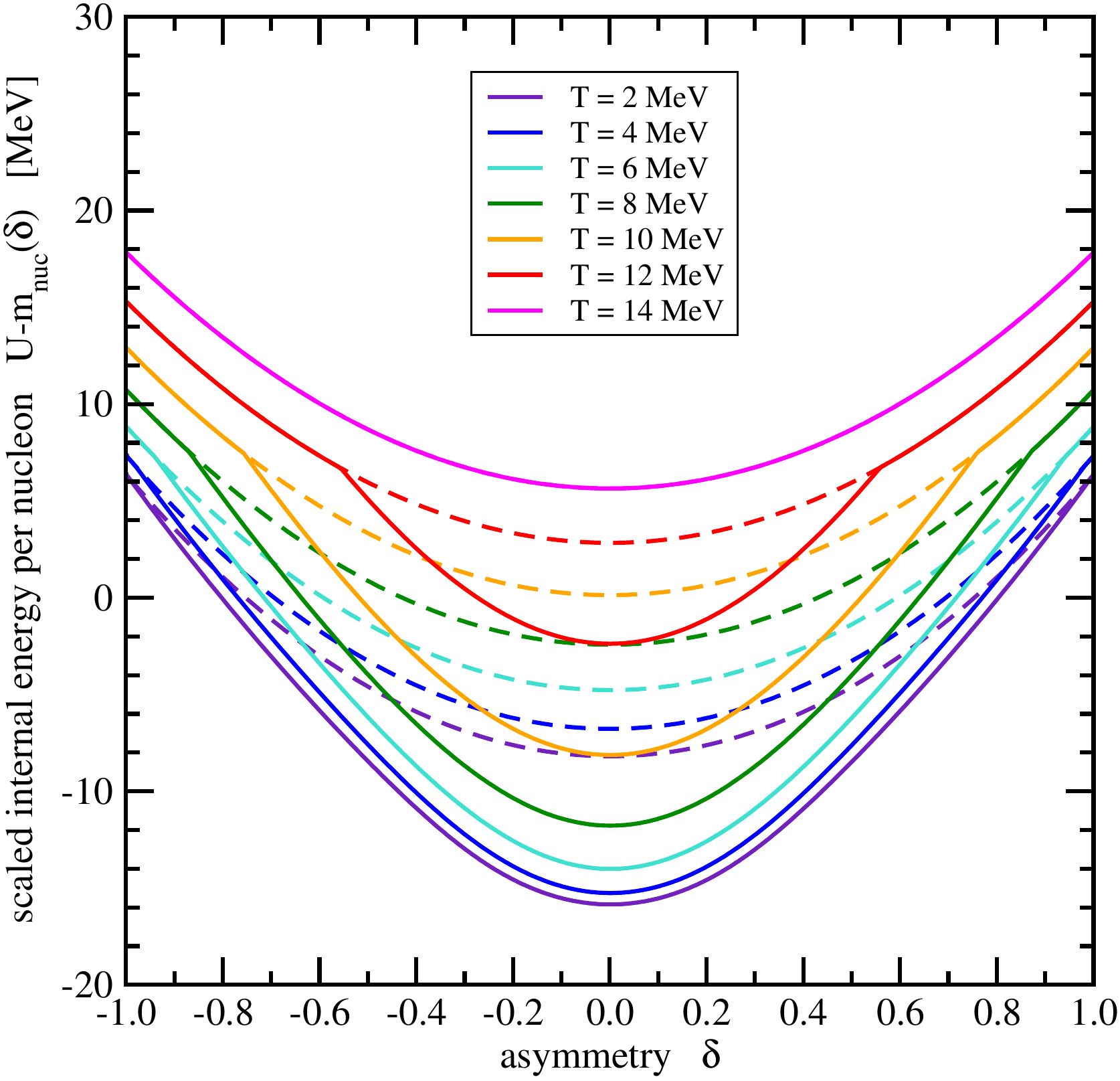}
\caption{\label{fig:U_alpha}%
Internal energy per nucleon $U$ corrected for the trivial mass
contribution $m_{\rm nuc}(\delta)$ as a function of the asymmetry
$\delta$ in nuclear matter at the critical baryon density $n_{\rm
  crit}$ of the DD2 paremetrization for different tempertures $T$
without (dashed lines) and with (full lines) liquid-gas phase
transition without cluster formation.}
\end{center}
\end{figure}

\subsubsection{Nuclear matter}

In nuclear matter, only nucleons are considered 
as constituent particles but not leptons. There is a liquid-gas phase 
transition at densities
below the nuclear saturation density and below the critical
temperature $T_{\rm crit}=13.724$~MeV of symmetric nuclear matter 
in the gRDF approach. 
In the present subsection no clusters
are considered in contrast to Ref.\ \cite{Typ10} because
the effects of the liquid-gas phase transition on the symmetry energy
are the main focus. The formation of clusters at low densities 
and low temperatures affects 
the main features of the phase transition 
only slightly since clusters appear in a substantial
amount only for densities, temperatures and asymmetries that lie
inside the coexistence region of the liquid-gas phase transition.

However, the liquid-gas phase transition will change
the dependence of the free energy per nucleon
\begin{equation}
 F(T,n_{B},\delta) = \frac{1}{n_{B}} f(T,n_{B},\delta)
\end{equation}
on the asymmetry $\delta$ for constant temperature $T$ and baryon
density $n_{B}$. The quantity
$f$ is the free energy density in the gRDF model, equation
(\ref{eq:f_dens}).
Similarly, the internal energy per nucleon is defined as
\begin{equation}
 U(T,n_{B},\delta) = \frac{1}{n_{B}} u(T,n_{B},\delta)
\end{equation}
with the internal energy density $u$ of equation (\ref{eq:u_dens}).

The dependence of $F$ and $U$ on $\delta$ for constant critical baryon
density
$n_{B}^{(\rm crit)} = 0.04515$~fm${}^{-3}$ is depicted in figures
\ref{fig:F_alpha} and \ref{fig:U_alpha}, respectively, for various
temperatures.
The trivial $\delta$ dependent contribution of the rest masses
\begin{equation}
\label{eq:m_nuc}
 m_{\rm nuc}(\delta) = \frac{1+\delta}{2} m_{n} + \frac{1-\delta}{2} m_{p}
\end{equation}
has been substracted in these figure for clarity. Dashed lines show
the results assuming uniform nuclear matter without a phase
transition.
There is a smooth variation of the energies, symmetric in $\delta$,
with an almost perfect quadratic dependence. With the
liquid-gas phase transition, we observe a reduction of the energies
that is most pronounced at symmetric nuclear matter. This reduction
is larger for lower temperatures and vanishes for $T \geq T_{\rm
  crit}$. Thus it is absent in the lines for $T = 14$~MeV. The free energy
per nucleon is a convex function of $\delta$ as required by
thermodynamical stability. But the internal
energy per nucleon $U$ exhibits a structure clearly indicating the transition 
to the region of coexisting phases at small asymmetries. The width of
this zone increases with decreasing temperature. It is also evident from
these two figures that the quadratic dependence on $\delta$ does not
hold for large $|\delta|$. There, it is closer to a linear dependence as
observed already for the mimimum curves in figure
\ref{fig:bea_nuc_alpha} for nuclei.

\begin{figure}[t]
\begin{center}
\includegraphics[width=8.5cm]{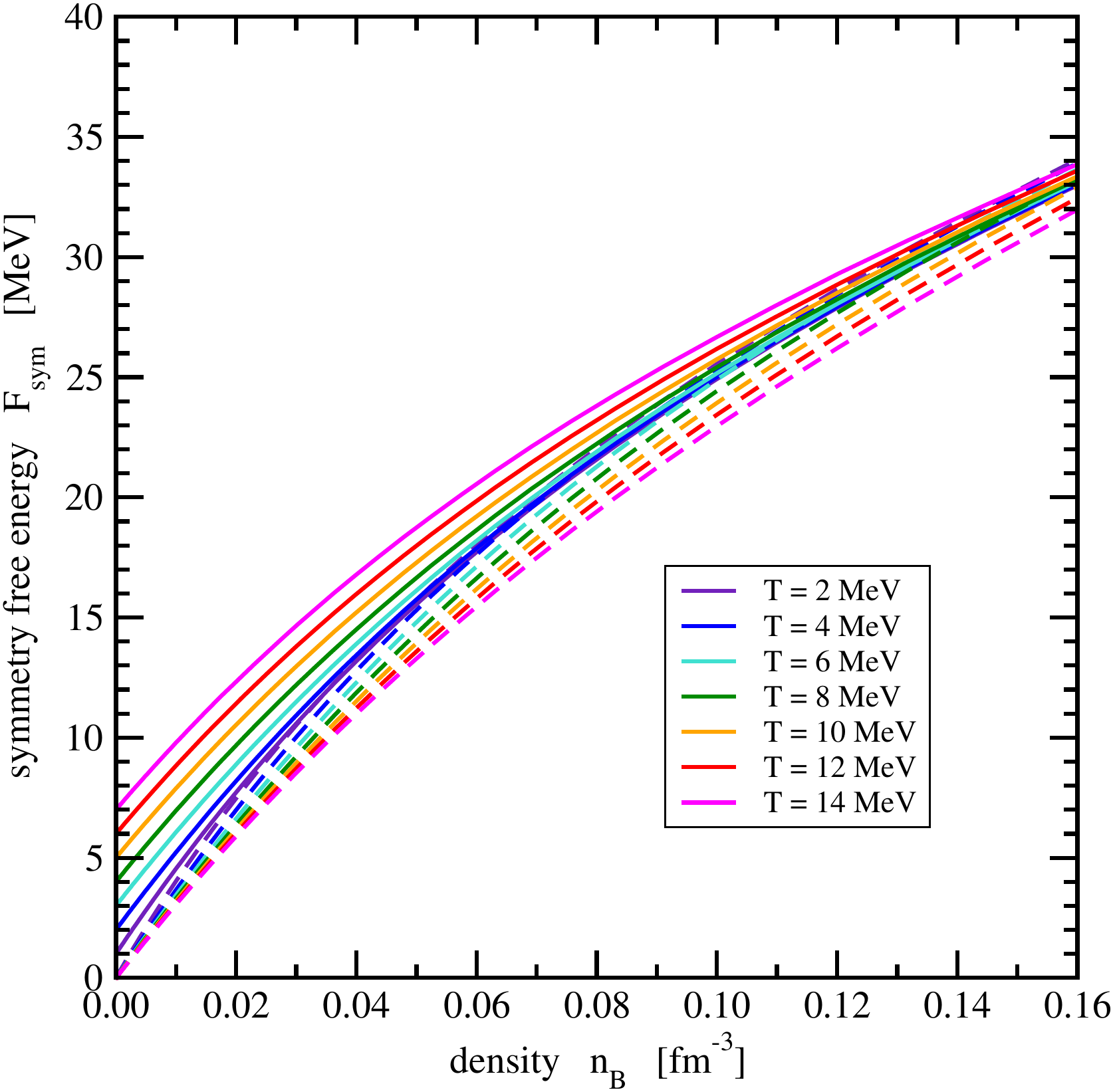}
\caption{\label{fig:fsym_nopt_lin}%
Symmetry free energy $F_{\rm sym}$ in uniform nuclear matter without
liquid-gas phase transition and without cluster formation 
as a function of the baryon density
$n_{B}$ for various temperatures in the second-derivative definition
(dashed lines) and the finite difference definition (full lines).}
\end{center}
\end{figure}

\begin{figure}[t]
\begin{center}
\includegraphics[width=8.5cm]{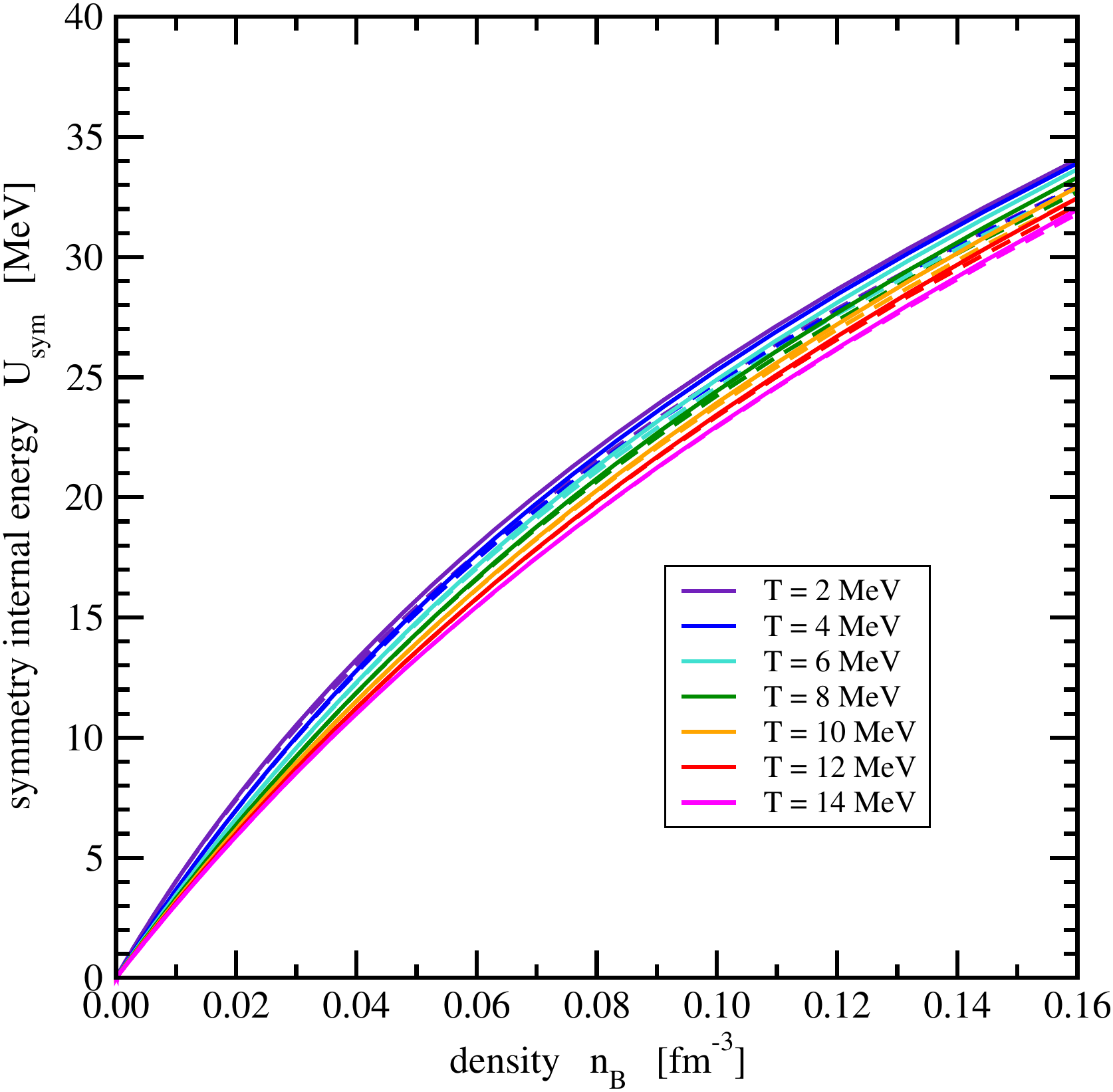}
\caption{\label{fig:esym_nopt_lin}%
Symmetry internal energy $U_{\rm sym}$ in uniform nuclear matter without
liquid-gas phase transition and without cluster formation
as a function of the baryon density
$n_{B}$ for various temperatures in the second-derivative definition
(dashed lines) and the finite difference definition (full lines).}
\end{center}
\end{figure}

As mentioned in subsection \ref{subsec:nuc_mat}, the symmetry free
energy $F_{\rm sym}(T,n_{B})$ can be defined by a second derivative as in equation
(\ref{eq:fsym_def})
or by a finite difference as 
\begin{eqnarray}
\label{eq:fsym_diff}
 \lefteqn{F_{\rm sym}(n_{B})}
 \\ \nonumber & = & \frac{1}{2} \left[ F(n_{B},+1)- 2F(n_{B},0) + F(n_{B},-1)\right]
\end{eqnarray}
Similarly, we have the symmetry internal energy
(\ref{eq:usym_def}) from a second derivative and the finite difference 
form
\begin{eqnarray}
\label{eq:usym_diff}
 \lefteqn{U_{\rm sym}(n_{B})}
 \\ \nonumber & = & \frac{1}{2} \left[ U(n_{B},+1)- 2U(n_{B},0) +
   U(n_{B},-1)\right] \: .
\end{eqnarray}
The differences between these definition are shown in figures
\ref{fig:fsym_nopt_lin} and \ref{fig:esym_nopt_lin} for uniform
nuclear matter without the liquid-gas phase transition.
The agreement of the two definitions are very good for the symmetry
internal energy $U_{\rm sym}$ at all densities. But for the symmetry
free energy larger systematic deviations are seen that can reach
several MeV. The symmetry free
energy $F_{\rm sym}(T,n_{B})$ approaches a finite value for $n_{B} \to
0$ that rises with the tempature $T$. It is due to the entropy
differences with 
\begin{equation}
 \lim_{n_{B}\to 0} F_{\rm sym}(T,n_{B}) = T \ln 2 \: .
\end{equation} 
In contrast to that, the symmetry
internal energy $U_{\rm sym}$ always approaches zero in this
limit.

\begin{figure}[t]
\begin{center}
\includegraphics[width=8.5cm]{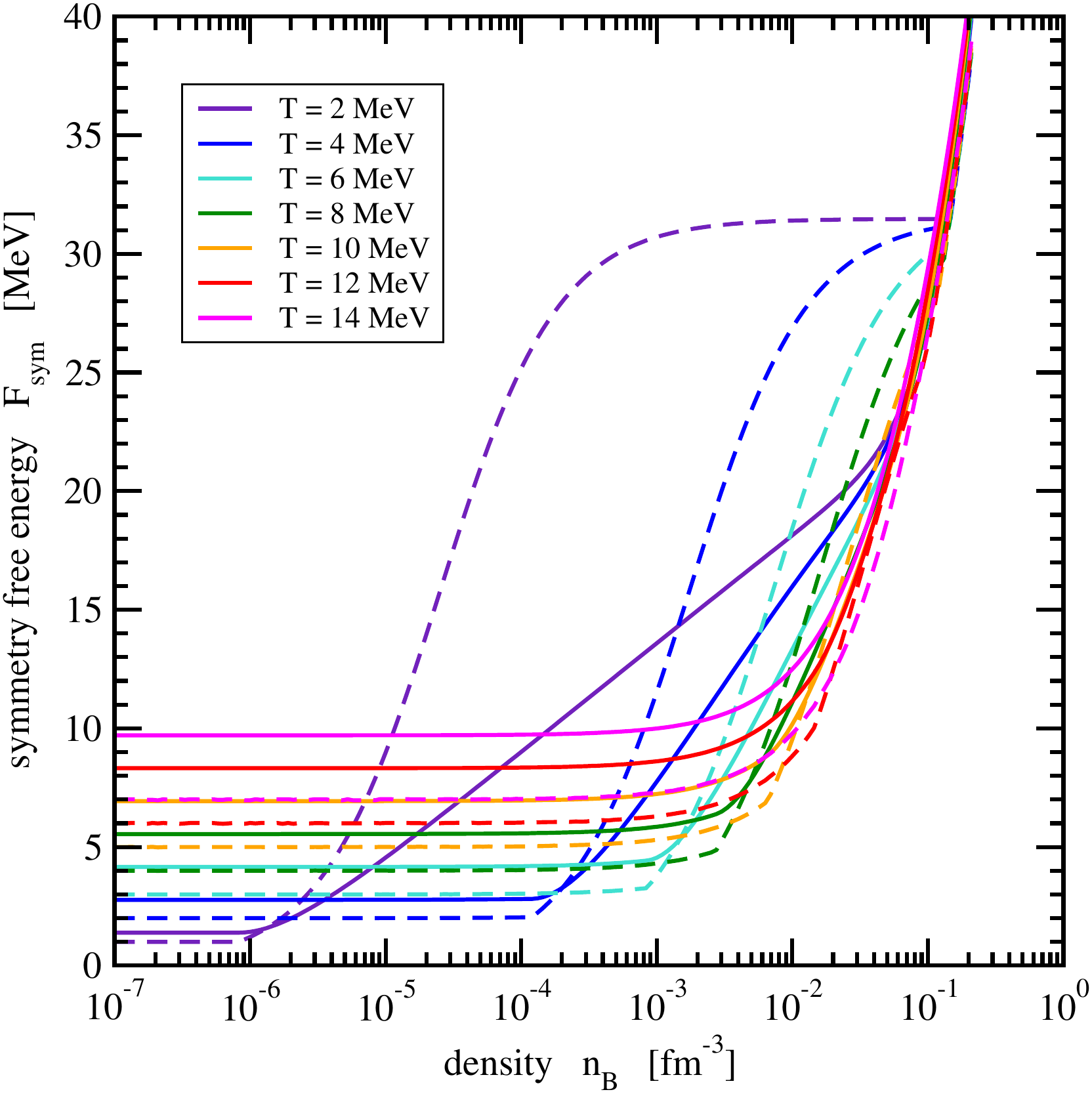}
\caption{\label{fig:fsym_pt_log}%
Symmetry free energy $F_{\rm sym}$ in nuclear matter with
liquid-gas phase transition but without cluster formation
as a function of the baryon density
$n_{B}$ for various temperatures in the second-derivative definition
(full lines) and the finite difference definition (dashed lines).}
\end{center}
\end{figure}

\begin{figure}[t]
\begin{center}
\includegraphics[width=8.5cm]{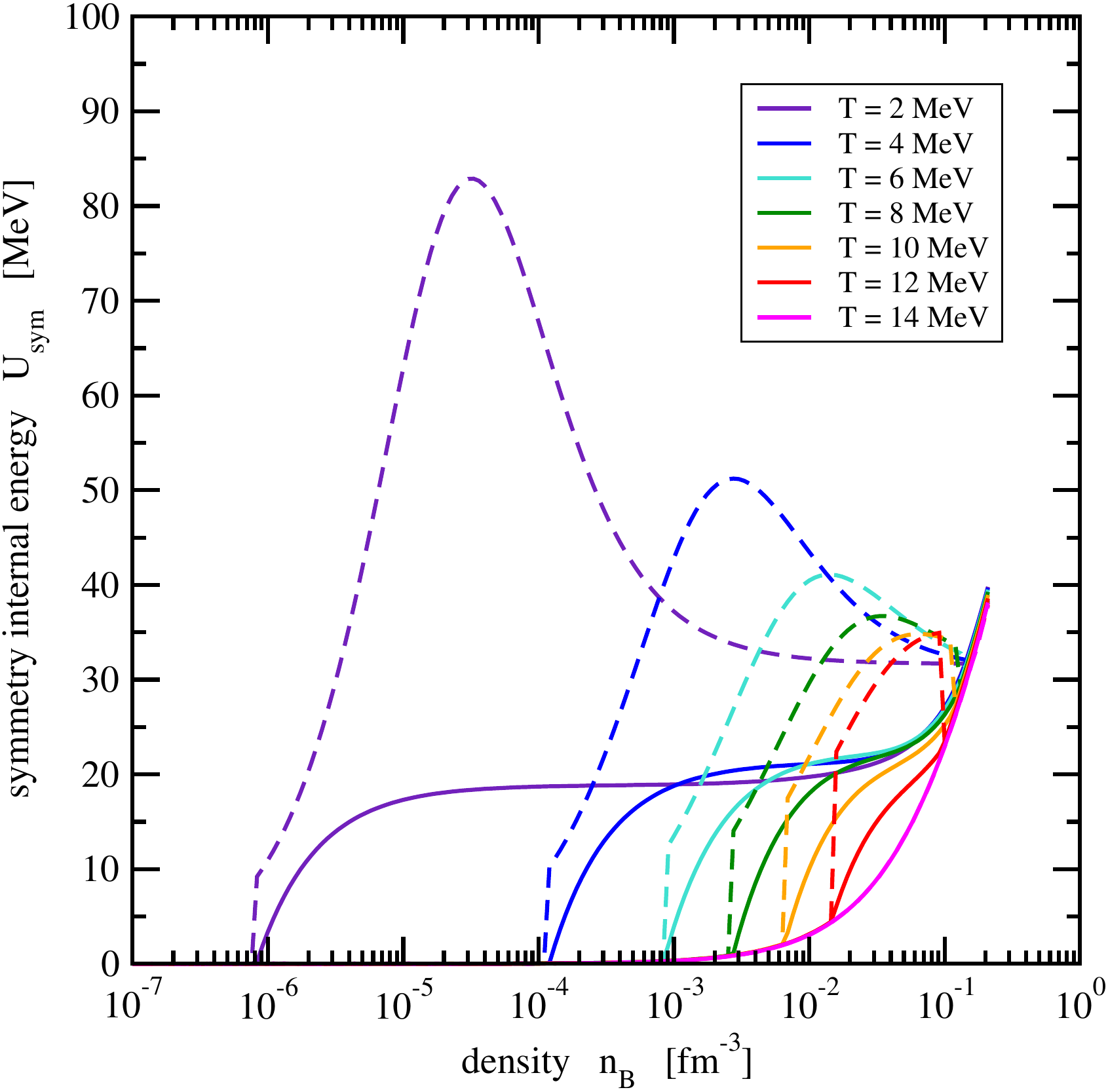}
\caption{\label{fig:esym_pt_log}%
Symmetry internal energy $U_{\rm sym}$ in nuclear matter with
liquid-gas phase transition but without cluster formation
as a function of the baryon density
$n_{B}$ for various temperatures in the second-derivative definition
(full lines) and the finite difference definition (dashed lines).}
\end{center}
\end{figure}

Now let us turn to the calculation with the liquid-gas phase
transition and perform the same comparison of the two definitions.
The corresponding results are depicted in figures
\ref{fig:fsym_pt_log} and \ref{fig:esym_pt_log} using logarithmic
scales on the axes for a better representation. A vast difference between
the two symmetry energy definitions is found in the region of the
coexisting phases. Only in the range of uniform nuclear matter the 
two approaches give similar results with systematically lower values of
the symmetry free energy when the derivative definition is used.
The finite difference formulas (\ref{eq:fsym_diff}) and
(\ref{eq:usym_diff}) give reasonable
quantitative results for the symmetry free energy and the symmetry internal energy
for all densities reflecting the difference in energies between
symmetric nuclear matter and pure neutron/proton matter. The second
derivative definitions (\ref{eq:fsym_def}) and (\ref{eq:usym_def})
however produce huge values of the
symmetry energy in the phase coexistence region, 
in particular at higher baryon densities close to the
transition to uniform nuclear matter. The finite difference
formula for the symmetry energy gives a better impression about the
variation of the energy per particle with the isospin variation.
Hence, we will use this definition in the following discussion.

\begin{figure}[t]
\begin{center}
\includegraphics[width=8.5cm]{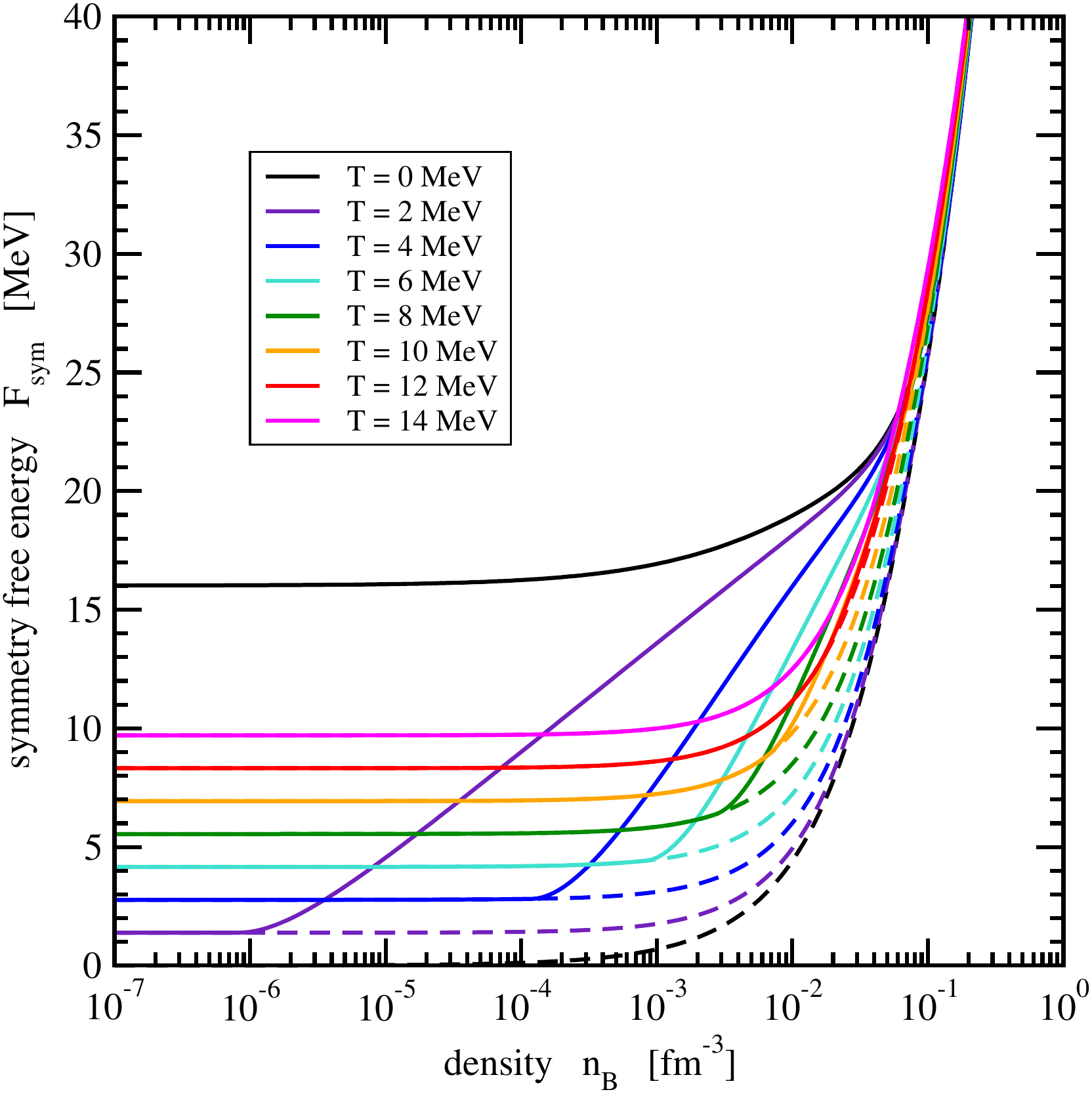}
\caption{\label{fig:fsym_log}%
Symmetry free energy $F_{\rm sym}$ in nuclear matter 
without cluster formation, without (dashed
lines) and with (full lines)
liquid-gas phase transition, as a function of the baryon density
$n_{B}$ for various temperatures using the finite difference
definition of the symmetry free energy.}
\end{center}
\end{figure}

\begin{figure}[t]
\begin{center}
\includegraphics[width=8.5cm]{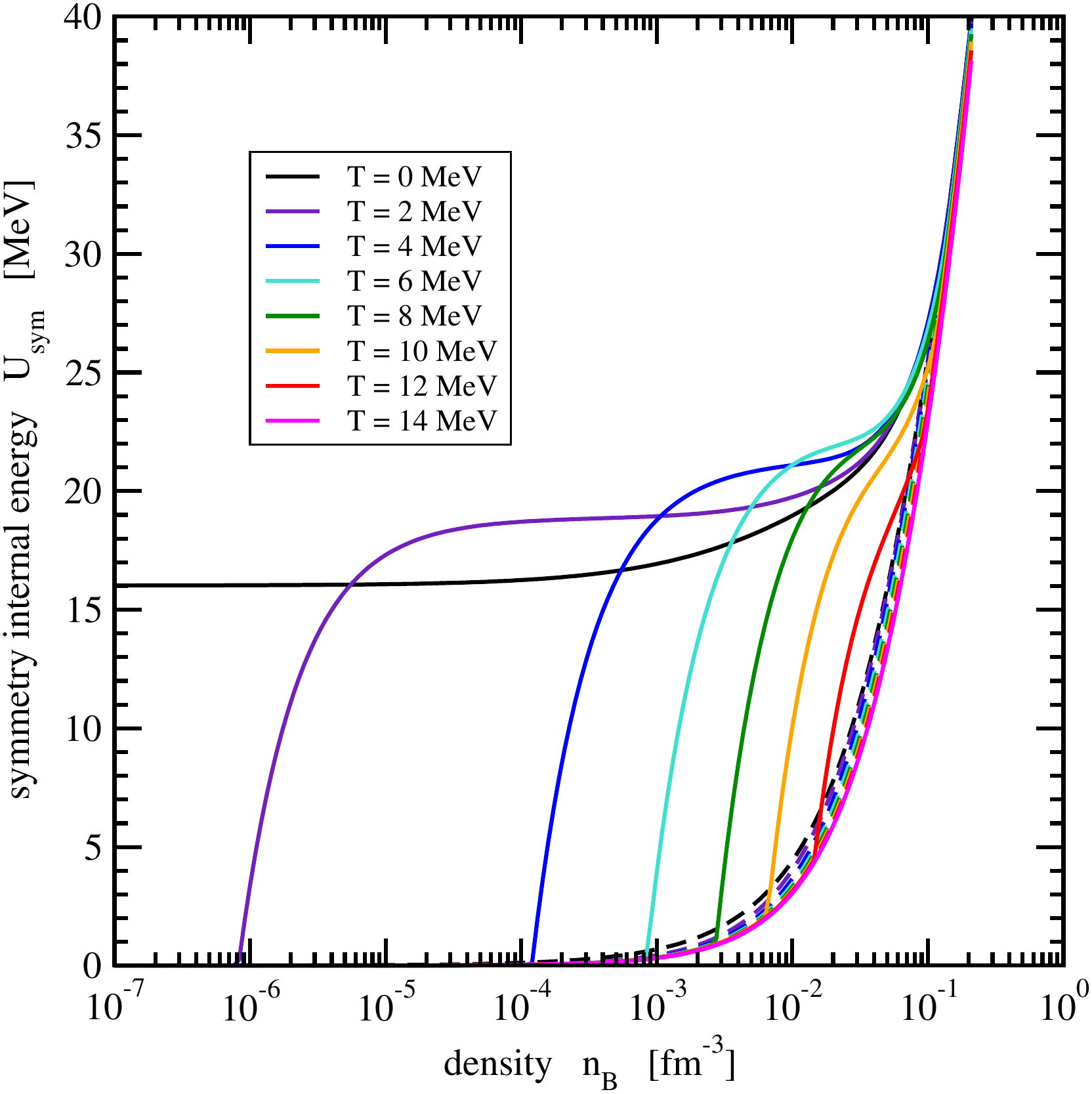}
\caption{\label{fig:esym_log}%
Symmetry internal energy $U_{\rm sym}$ in nuclear matter 
without cluster formation, without (dashed
lines) and with (full lines)
liquid-gas phase transition, as a function of the baryon density
$n_{B}$ for various temperatures using the finite difference
definition of the symmetry internal energy.}
\end{center}
\end{figure}

In figures \ref{fig:fsym_log} and \ref{fig:esym_log}, the symmetry free
energy and the symmetry internal energy without and with liquid-gas
phase transition are depicted in a linear scale for the symmetry
energies employing the finite difference formula. The effect of the
phase transition is easily discerned. We emphasize that
the values at low densities are a result of the separation of phases
and not due to cluster formation that is not taken into account in the
nuclear matter calculations.
The effects of clusters will be considered only in
the next subsection. Due to the different low-density
limits, the effect on the symmetry internal energy is more pronounced.
A particular interesting case is the low-density behavior of the
symmetry energies for zero temperature that is depicted in these
figure, too. Both the symmetry free and
symmetry internal energy approach a finite value in this exceptional
situation with $n_{B}\to 0$. The limiting value is just the binding energy $B_{\rm sat}
\approx 16$~MeV of nuclear matter at saturation.

\subsubsection{Stellar matter}

\begin{figure}[t]
\begin{center}
\includegraphics[width=8.5cm]{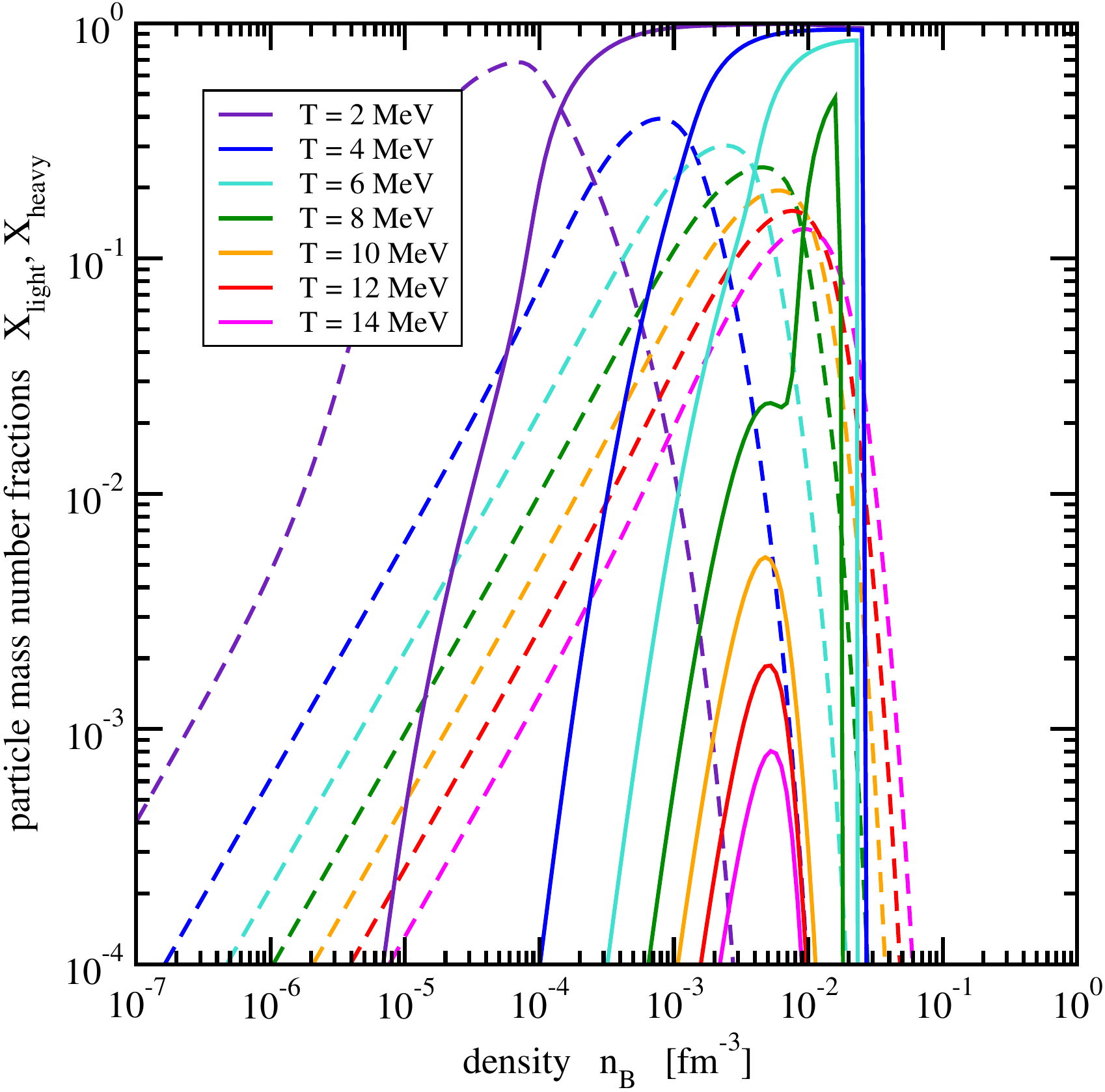}
\caption{\label{fig:frac_all}%
Mass number fractions of light (dashed lines) and heavy nuclei (full
lines) in stellar matter with asymmetry $\delta = 0$ for different
temperature $T$ as a function of the baryon number density $n_{B}$.}
\end{center}
\end{figure}

\begin{figure}[t]
\begin{center}
\includegraphics[width=8.5cm]{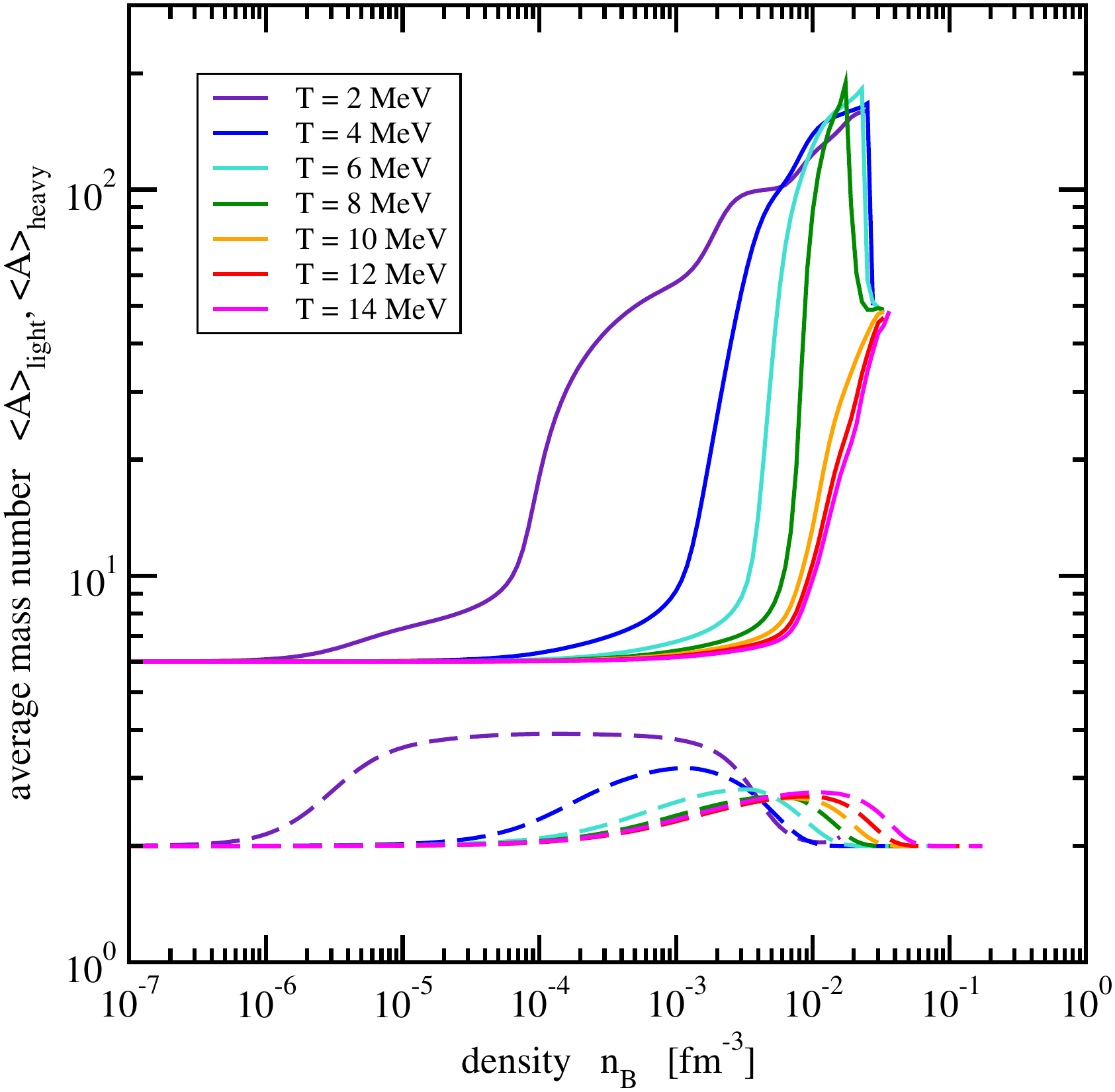}
\caption{\label{fig:A_all}%
Average mass number of light (dashed lines) and heavy nuclei (full
lines) in stellar matter with asymmetry $\delta = 0$ for different
temperature $T$ as a function of the baryon number density $n_{B}$.}
\end{center}
\end{figure}

In the calculcation of stellar matter properties the full set of
constituents in the gRDF model is used, i.e. nucleons, electrons,
and all nuclei with $A\leq 350$. In order to follow the evolution of
the chemical composition, we introduce the particle fractions
\begin{equation} 
 X_{\rm light} = \frac{1}{n_{B}} \sum_{i\in \mathcal{S}_{\rm light}}
   A_{i} n_{i}
\end{equation}
of the light clusters (set $\mathcal{S}_{\rm light}= \left\{ 
{}^{2}\mbox{H}, {}^{3}\mbox{H}, {}^{3}\mbox{He}, {}^{4}\mbox{He}\right\}$) and 
\begin{equation} 
 X_{\rm heavy} = \frac{1}{n_{B}} \sum_{i\in \mathcal{S}_{\rm heavy}}
   A_{i} n_{i}
\end{equation}
of the heavy clusters
(set $\mathcal{S}_{\rm heavy}=\left\{ (N_{i},Z_{i}) | A_{i} > 4 \right\}$). 
In figure \ref{fig:frac_all} the quantities
$X_{\rm light}$ and $X_{\rm heavy}$ are shown as a function of the
baryon number density $n_{B}$ for various temperatures in stellar
matter with asymmetry $\delta = 0$. At low densities, light clusters
are the prevailing species. Heavy clusters dominate the composition
at higher densities as long as the temperature is not too high.
When the density approaches nuclear saturation density, all cluster
dissolve as expected. Thus the model accounts for the Mott effect.
It is due to the mass shifts of clusters in the
gRDF model, which are given in appendix
\ref{sec:mshifts} including the electron screening in the Wigner-Seitz
approximation. It is a more microscopic 
alternative to the often used excluded volume mechanism to suppress 
the occurrence of nuclei in a dense medium.
The effective degeneracy factors accounting for internal excitations
are given in appendix \ref{sec:geff}. 

More information on the chemical composition is given by the average
mass numbers
\begin{equation}
 \langle A \rangle_{\rm light} = 
 \frac{\sum_{i \in \mathcal{S}_{\rm light}} A_{i}n_{i}}{\sum_{i \in \mathcal{S}_{\rm light}}n_{i}}
\end{equation}
and
\begin{equation}
 \langle A \rangle_{\rm heavy} = 
 \frac{\sum_{i \in \mathcal{S}_{\rm heavy}} A_{i}n_{i}}{\sum_{i \in \mathcal{S}_{\rm heavy}}n_{i}}
\end{equation}
of the light and of the heavy component
that are depicted in figure \ref{fig:A_all}. At very low densities, the
cluster composition is mainly given by light nuclei, i.e.\ ${}^{2}$H in
the light and ${}^{6}$Li in the heavy component. With increasing
density, 
the average mass number rises. The light clusters turn into $\alpha$
particles and heavy cluster become substantially more massive but 
$\langle A \rangle_{\rm heavy}$ does not exceed $200$ in the present
gRDF model for $\delta = 0$.
At low temperatures, shell effects in the nuclear binding energies 
cause the particular structure of the average mass number and charge
number evolution
with the density. At higher temperatures, these effects are washed out.
The density range where clusters give an important contribution to the
chemical composition shrinks with increasing temperature. Beyond
$T\approx 10$~MeV, heavy clusters quickly disappear.

\begin{figure}[t]
\begin{center}
\includegraphics[width=8.5cm]{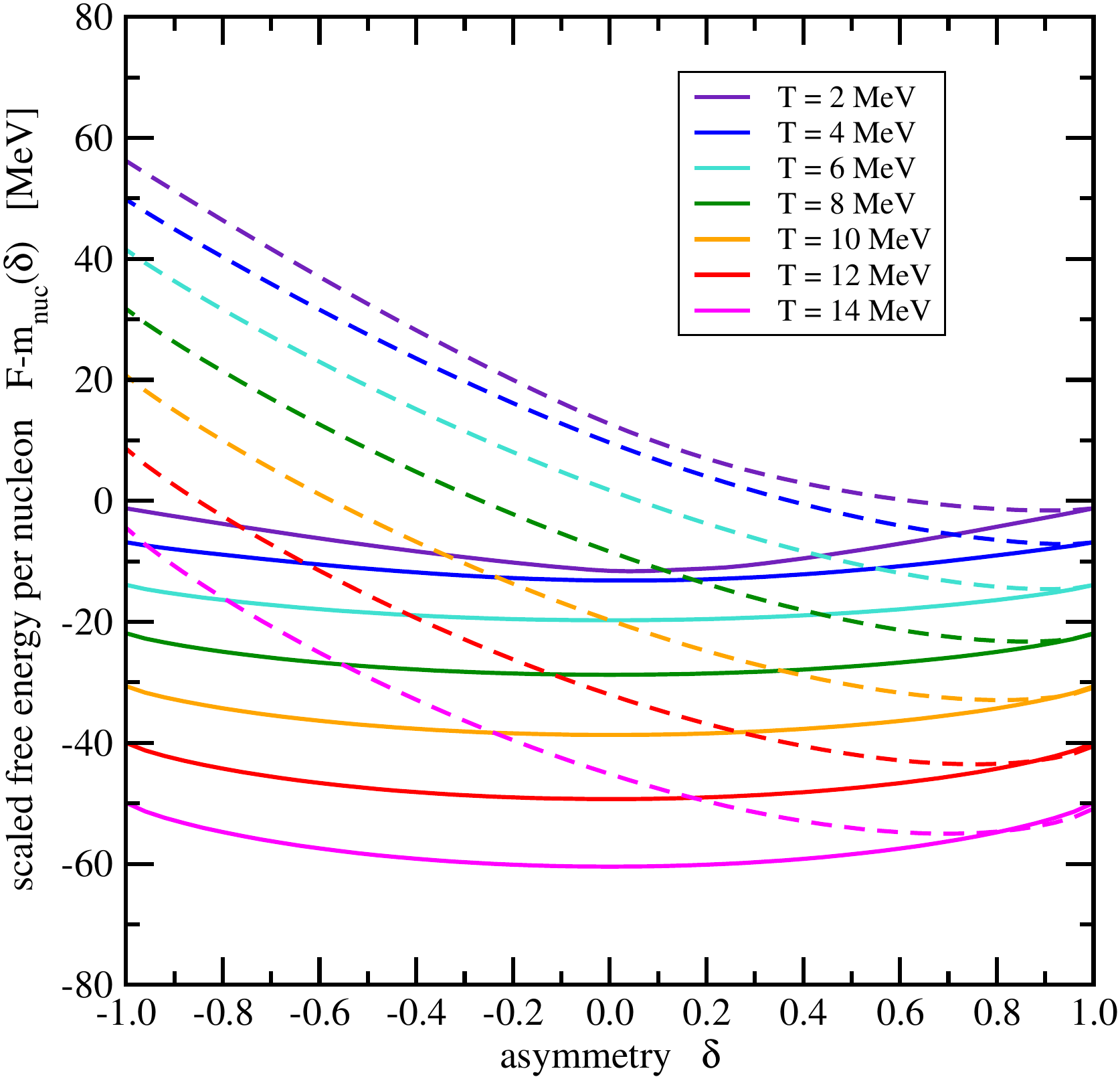}
\caption{\label{fig:f_all}%
Free energy per nucleon $F$ corrected for the trivial mass
contribution $m_{\rm nuc}(\delta)$ as a function of the asymmetry
$\delta$ in stellar matter at a baryon density of
$n_{B}=0.002$~fm$^{-3}$ for different temparatures $T$ without
(dashed lines) and with (full lines) the Coulomb correction.}
\end{center}
\end{figure}

\begin{figure}[t]
\begin{center}
\includegraphics[width=8.5cm]{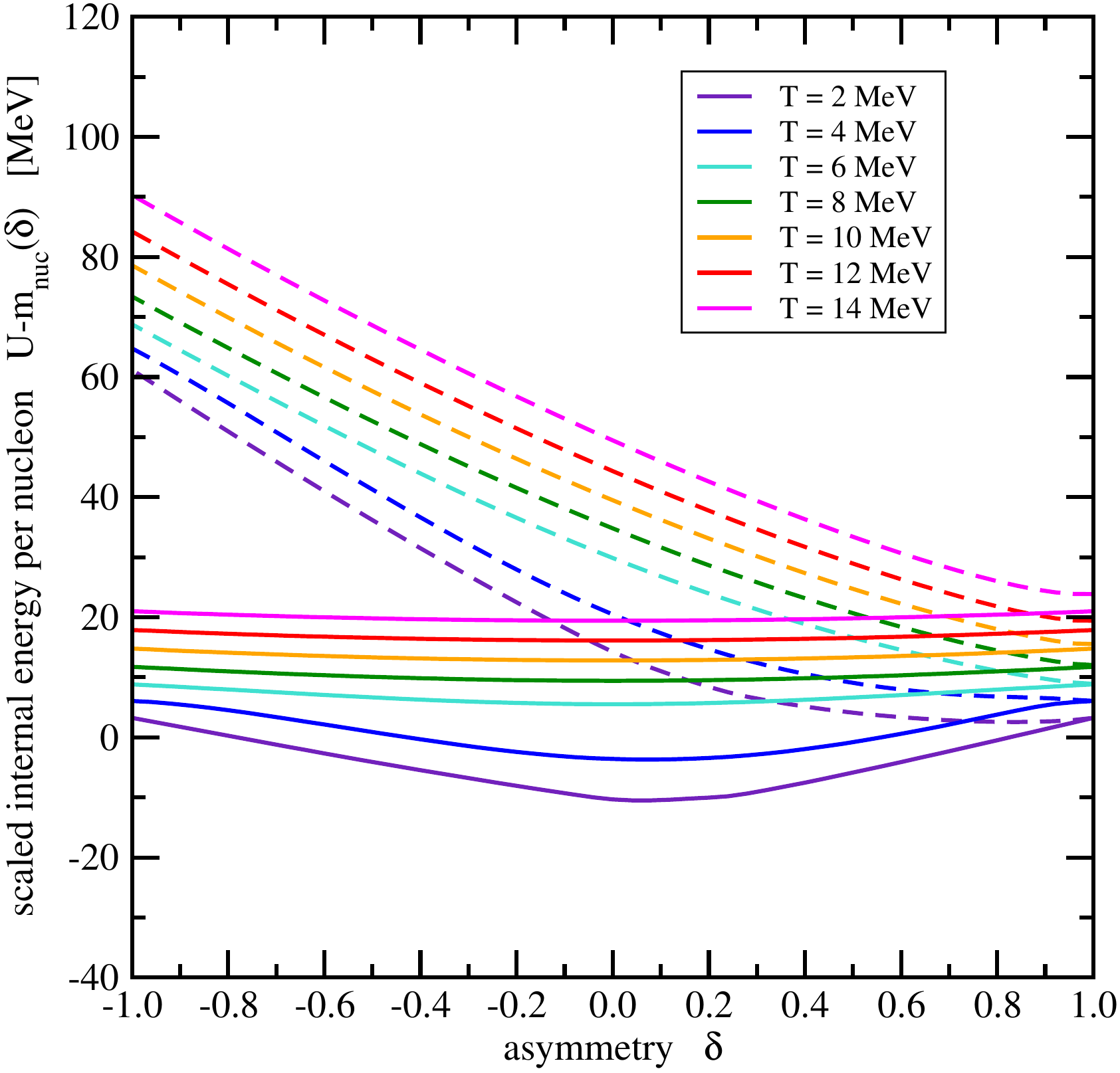}
\caption{\label{fig:e_all}%
Internal energy per nucleon $U$ corrected for the trivial mass
contribution $m_{\rm nuc}(\delta)$ as a function of the asymmetry
$\delta$ in stellar matter at a baryon density of
$n_{B}=0.002$~fm$^{-3}$ for different temparatures $T$ without
(dashed lines) and with (full lines) the Coulomb correction.}
\end{center}
\end{figure}

Electrons are an essential component in stellar matter because they
guarantee the charge neutrality of the system and screen the Coulomb
interaction at high densities. They contribute in a sizeable amount to
the thermodynamic quantities, in particular to the energies and
pressure. Because the electronic contribution is not symmetric in the
asymmetry parameter $\delta$, the isospin symmetry as it appears in
nuclear matter does not hold any more. This is clearly seen in figures
\ref{fig:f_all} and \ref{fig:e_all} which 
show with dashed lines the variation of the free
energy $F$ and internal energy $U$ per nucleon with
$\delta$ for stellar matter at a constant baryon density 
$n_{B} = 0.002$~fm${}^{-3}$. The contribution of the electrons leads to
an increase of the energies in proton rich matter.

Similarly as in the Bethe-Weizs\"{a}cker formula (\ref{eq:BW}) the
effect of the Coulomb interaction and the electronic contribution
has to be removed from the energies (apart from the trivial
neutron-proton mass difference effect) in order to obtain values for
the symmetry energies that are comparable to those of nuclei or
nuclear matter. For this purpose, we substract the contribution of the
electrons from the total thermodynamical quantities. This is easily
carried out in the formulation of the gRDF model in subsection
\ref{subsec:gRDF_thermo}. The second correction concerns the Coulomb
contribution to the binding energies of the nuclei. In the medium, the
Coulomb energy is already partially screened as described by the 
energy shift $\Delta E_{i}^{(\rm Coul)}$ in equation
(\ref{eq:DE_Coul}).
Thus we have to add only the remaining Coulomb shift that is required 
to obtain binding energies of nuclei without the Coulomb interaction.
This is performed in the calculation for all nuclei with the proper
weights proportional to their densities. The energies modified in this 
manner are plotted in figures \ref{fig:f_all} and 
\ref{fig:e_all} with full lines. We observe an almost perfect
symmetry with respect to $\delta=0$. The effect of the cluster
formation in the medium on the shape becomes more obvious after the
electron and Coulomb corrections were considered. For low
temperatures there is a considerable reduction of the energies for small
values of $\delta$. For temperatures above $T \approx 6$~MeV the effect is
almost disappearing as expected from the information on the chemical
composition in figures \ref{fig:frac_all} and \ref{fig:A_all}
as well as from the occurrence of the phase transition in 
figures \ref{fig:fsym_log} and \ref{fig:esym_log}.
The full lines in figures \ref{fig:f_all} and \ref{fig:e_all}
exhibit the same trend that was already depicted in figure 12 of Ref.\
\cite{Typ10} where only light clusters were included in the
theoretical model. Due to the inclusion of the heavy nuclei in the
current calculation, the minimum is rounded and less triangular shaped.

\begin{figure}[t]
\begin{center}
\includegraphics[width=8.5cm]{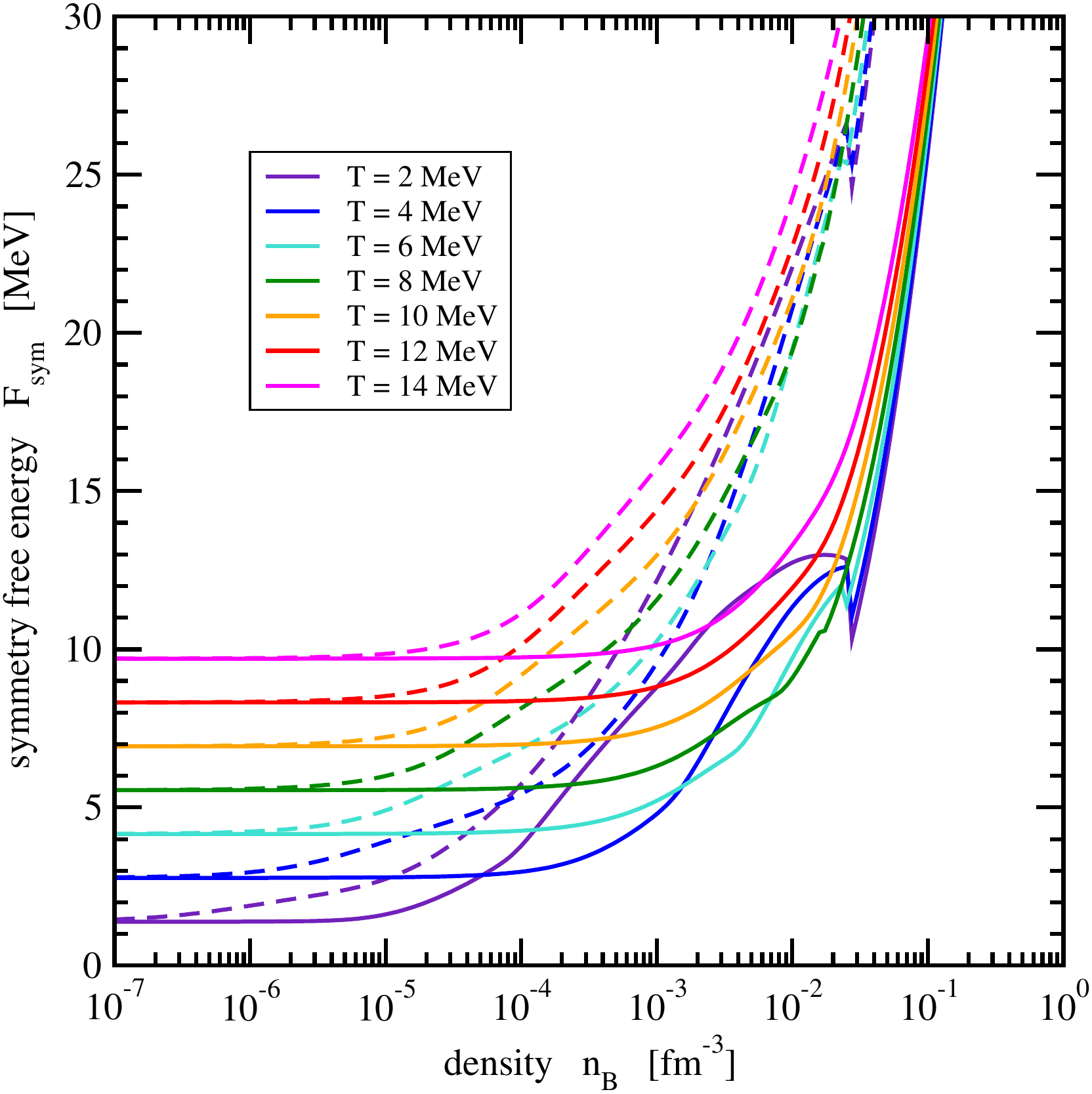}
\caption{\label{fig:fsym_all}%
Symmetry free energy $F_{\rm sym}$ in stellar matter without (dashed
lines) and with (full lines)
Coulomb and electron correction as a function of the baryon density
$n_{B}$ for various temperatures using the finite difference
definition of the symmetry free energy.}
\end{center}
\end{figure}

\begin{figure}[t]
\begin{center}
\includegraphics[width=8.5cm]{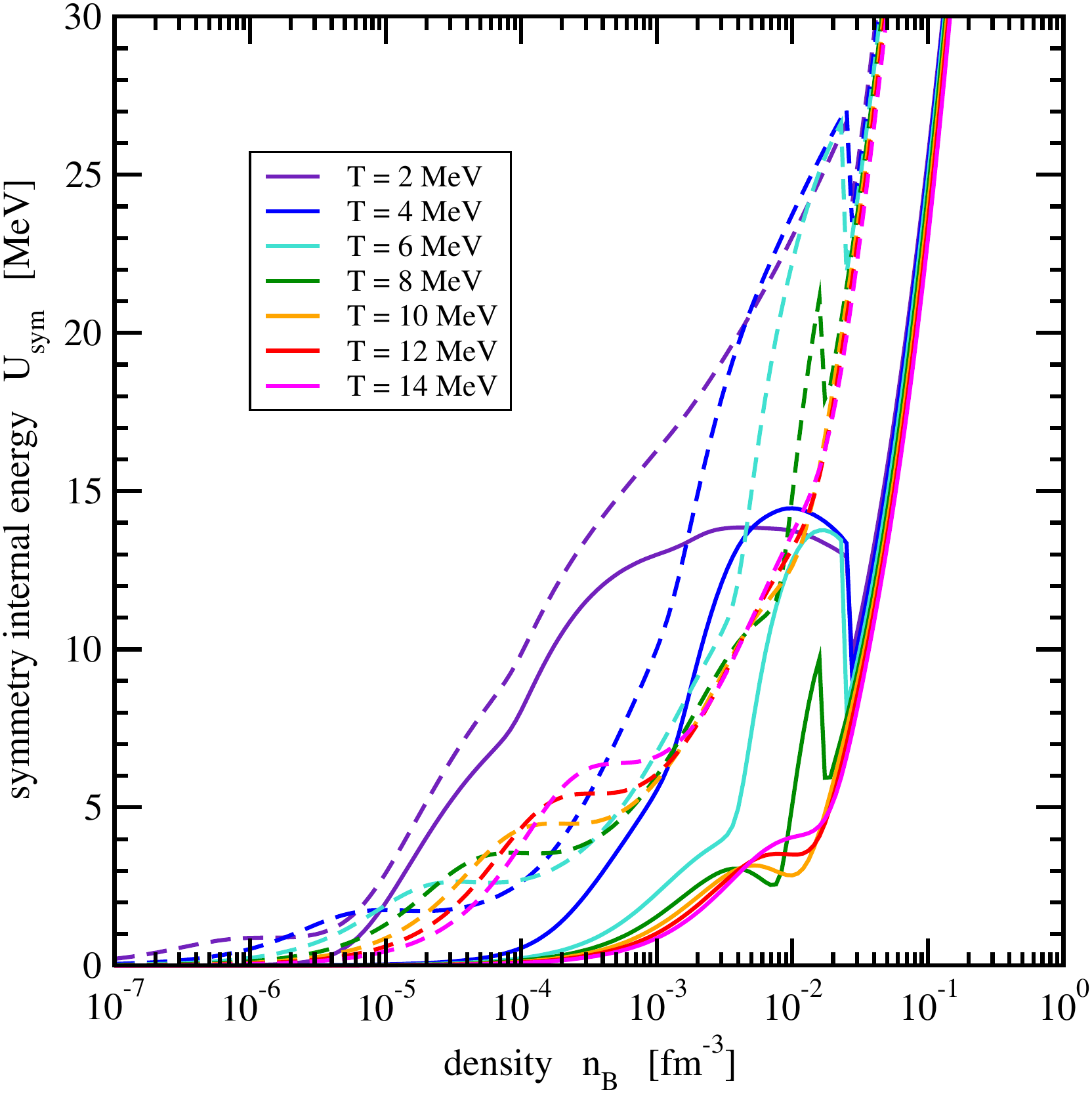}
\caption{\label{fig:esym_all}%
Symmetry internal energy $U_{\rm sym}$ in stellar matter without (dashed
lines) and with (full lines)
Coulomb and electron correction as a function of the baryon density
$n_{B}$ for various temperatures using the finite difference
definition of the symmetry internal energy.}
\end{center}
\end{figure}

Using the finite-difference definition of the symmetry energy, the
dependence of the symmetry free and symmetry internal energies as a
function of the baryon density are obtained without and with the
electron and Coulomb corrections. The result are displayed in figures
\ref{fig:fsym_all} and \ref{fig:esym_all}. The differences between the
uncorrected and corrected results are clearly visible. With the
correction, the symmetry energies are too large because of the
larger variation of the energies with $\delta$ for constant baryon
density due to the electronic contribution. Only at low densities both
calculations will merge since the charge densities and Coulomb shifts
become very small. We also notice that the clustering in dense stellar
matter leads to reduced symmetries as compared to those in nuclear
matter with liquid-gas phase transition. This is true both in absolute
value and in the extension of the density range. 
Comparing the results in figures \ref{fig:fsym_all} and
\ref{fig:esym_all}
with those in figure 13~(a) and 14~(a)
of Ref.\ \cite{Typ10} we observe that the occurrence of heavy nuclei 
increases the symmetry energies at low temperatures due to their
larger binding energies as those of the light clusters.

Cluster formation was not included
in the calculation of the symmetry energy in nuclear
matter with liquid-gas phase transition.
In contrast, no phase transition construction was applied to the
presented results for stellar
matter in order to expose the effect of clustering as clearly as
possible. As a consequence, some non-monotonic behavior of the curves in
figures \ref{fig:fsym_all} and \ref{fig:esym_all} is observed 
at lower temperatures. This is caused by
the sudden disappearance of the clusters with increasing density.
The effect would vanish when the phase transition was fully accounted
for. The phase transition construction in stellar matter is
somewhat different as in nuclear matter due to the additional
conserved charge, the lepton number, and the charge neutrality
condition. See appendix \ref{sec:lgpt} for more details. 
However, at the relevant temperatures and densities, a more refined
calculation should take into account the appearance of ''pasta''
structures with complicated spatial density distributions that
smoothen the transition from matter with clusters to homogeneous
matter. A liquid-gas type phase transition construction can only
roughly represent this transition.
We leave the full treatment to a future publication in the context of
providing a global equation of state table of stellar matter
for astrophysical applications.


\section{Conclusions}
\label{sec:concl}

The symmetry energy is a valuable concept to characterize the
dependence of the energy on the isospin asymmetry of a system. 
However, due to the different definitions of this quantity
and the specific thermodynamic conditions of the system, a comparison
of the symmetry energies derived from different sources needs a careful
consideration of possible discrepancies. Besides finite nuclei,
the symmetry energy of dense matter as a function of the density
is of particular interest in theoretical and experimental investigations.

In the present work, the symmetry free energy and symmetry internal
energy of nuclear matter and of stellar matter were extracted from 
theoretical calculations employing a generalized relativistic density
functional approach, which allows to describe the apperance of
the liquid-gas phase transition or the formation and
dissolution of finite-size clusters. In stellar matter, corrections
for the existence of electrons and for the action of the
electromagnetic interaction are required in order to
extract the pure nuclear symmetry energy.

Results of various definitions for the symmetry energy were
presented. In systems with phase transition or cluster degrees of
freedom, substantial differences for the symmetry energy are found
by comparing the definitions using second derivatives or finite
differences. The latter approach, which compares symmetric matter with
pure neutron and proton matter, seems to give more reasonable sizes of
the symmetry energy. The occurrence of spatially inhomogeneous density
distributions causes an increase of the symmetry energies at low
densities, in particular at low temperatures. This is in strong
contrast to theoretical calculations assuming uniform uncorrelated
matter.

The comparison of experimentally determined symmetry energies with
those extracted from theoretical calculations is neither straightforward
nor necessarily direct. This topic deserves a more extended
discussion but it is beyond the scope of the present paper.
However, a few remarks are in order. In most investigations of heavy-ion
collisions, theoretical model simulations are utilized in order to
describe experimental observables that are sensitive to the isospin
asymmetry. As a result the isovector dependence of the interaction
of the underlying model or energy density functional 
is explored. It is connected to the density dependence of the
symmetry energy for homogeneous Coulomb-less nuclear matter, which
is conveniently encoded in quantities like $J$,
$L$, or $K_{\rm sym}$ for an easy comparison of models. This does not
mean that clusterization effects are not accounted for in the
experiments or that the extracted values of the  coefficients $J$,
$L$, or $K_{\rm sym}$ represent the physical symmetry energy.
Only in few cases, e.g.\ in Ref.\ \cite{Nat10}, it is attempted to
extract a symmetry
energy directly without the need to consider an intermediate simulation with a
theoretical model. In these cases, effects of clusterization as they
appear in the physical system under study will naturally affect the
extracted symmetry energy. The considerations of the present paper
apply to the comparison of symmetry energies that were directly determined
in experiments to those of indirect approaches employing, e.g.,
numerical simulations of heavy-ion collisions.

\begin{acknowledgement}
This work was supported by
by the Helmholtz Association (HGF) through the Nuclear Astrophysics
Virtual Institute (VH-VI-417),
by CompStar, a Research Networking Program
of the European Science Foundation (ESF),
by CompStar-POL
and by the Helmholtz International
Center for FAIR within the framework of the LOEWE program launched
by the state of Hesse via the Technical University Darmstadt.
D.B.\ was supported by NCN within the ``Maestro'' program
under grant No.\ DEC-2011/02/A/ST2/00306 and by RFBR under grant No.
11-02-01538-a.
\end{acknowledgement}

\appendix

\section{Liquid-gas phase transition in dense matter}
\label{sec:lgpt}

\begin{figure}[t]
\begin{center}
\includegraphics[width=8.5cm]{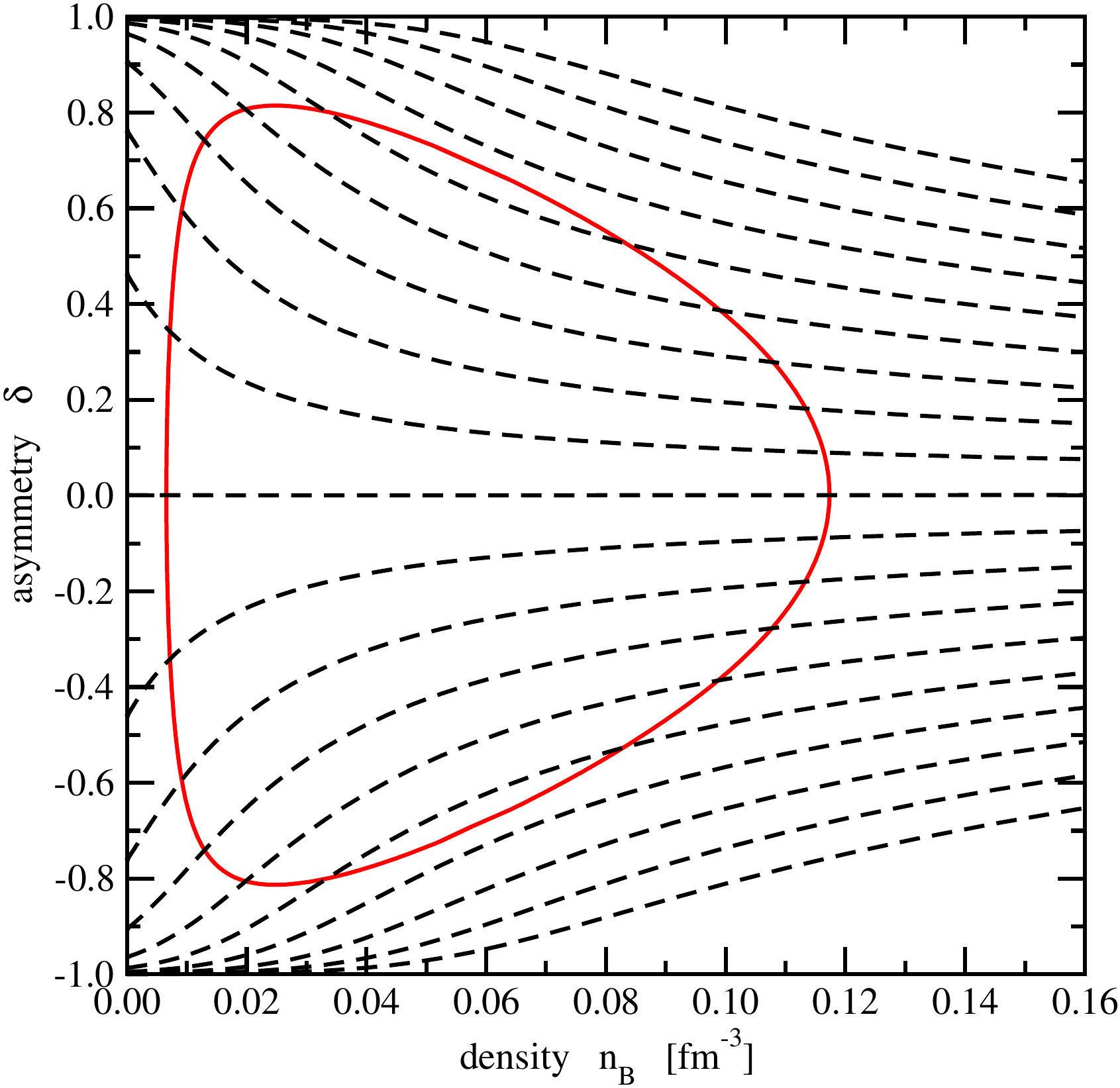}
\caption{\label{fig:bino_T10}%
Binodal (full red line)
and lines of equal charge chemical potential $\mu_{Q}$ (dashed black
lines, in steps of $10$~MeV) for a temperature of $T=10$~MeV 
in the gRDF model of nuclear matter without cluster formation
in the asymmetry-density plane.}
\end{center}
\end{figure}

The thermodynamic state of dense matter is completely determined
when the independent variables, i.e.\ the temperature and the
densities of the conserved charges, are chosen. 
It can be found by a global minimization of the free energy
density $f$. In some regions of the parameter space, dense matter will
separate into coexisting phases. The cases of nuclear matter and
stellar matter have to be distinguished for the construction of the
phase transition, which is briefly presented in the
following. A more detailed discussion of the phase transition
construction with the isospin degree of freedom and the reduction from
a general Gibbs construction to the simpler Maxwell construction can be found in
Ref.\ \cite{Duc06}.

\subsection{Nuclear matter}

In this case, the free energy density is a function of the 
independent variables $T$, $n_{B}$, and $n_{Q}$.
For parameters $n_{B}$ and $n_{Q} = (1-\delta)
n_{B}/2$ inside the binodals shown in figure \ref{fig:binodals_log}
the correct free energy density
is found by a linear interpolation of the
free energy densities of the coexisting phases 
$\pi = 1$ and $\pi = 2$ at two points that lie on the binodal 
of given temperature $T$ and have identical intensive variables, i.e.\
pressure
\begin{equation}
 p^{(\rm coex)} = p(T,n_{B}^{(1)},n_{Q}^{(1)}) =
 p(T,n_{B}^{(2)},n_{Q}^{(2)}) \: ,
\end{equation}
baryonic chemical potential 
\begin{equation}
 \mu_{B}^{(\rm coex)} = 
 \mu_{B}(T,n_{B}^{(1)},n_{Q}^{(1)}) =
 \mu_{B}(T,n_{B}^{(2)},n_{Q}^{(2)}) \: ,
\end{equation}
and charge chemical potential
\begin{equation}
 \mu_{Q}^{(\rm coex)} = 
 \mu_{Q}(T,n_{B}^{(1)},n_{Q}^{(1)}) =
 \mu_{Q}(T,n_{B}^{(2)},n_{Q}^{(2)}) \: .
\end{equation}
A practical way of finding the correct baryon ($n_{B}^{(1)}$, $n_{B}^{(2)}$) 
and charge ($n_{Q}^{(1)}$, $n_{Q}^{(2)}$) densities of the coexisting
phases uses the modified thermodynamic potential
\begin{equation}
\label{eq:def_ftilde}
 \tilde{f}_{\rm nuc}(T,n_{B},\mu_{Q}) = f(T,n_{B},n_{Q}) - \mu_{Q} n_{Q}
\end{equation}
that depends only on a single extensive-like variable, here the density
$n_{B}$, apart from further intensive variables. The densities
$n_{B}^{(1)}$ and $n_{B}^{(2)}$ are then found by a simple
one-dimensional phase transition construction, e.g.\ the usual Maxwell
construction, requiring 
\begin{eqnarray}
 p^{(\rm coex)} & = & p(T,n_{B}^{(1)},\mu_{Q}) = p(T,n_{B}^{(2)},\mu_{Q})
 \\ \nonumber & = &
 n_{B}^{2} \left. 
 \frac{\partial \left(\tilde{f}_{\rm nuc}/n_{B}\right)}{\partial n_{B}}
\right|_{T,\mu_{Q},n_{B} = n_{B}^{(\pi)}}
\end{eqnarray}
and
\begin{eqnarray}
 \mu_{B}^{(\rm coex)} & = &
 \mu_{B}(T,n_{B}^{(1)},\mu_{Q}) = \mu_{B}(T,n_{B}^{(2)},\mu_{Q}) 
 \\ \nonumber & = & 
 \left. 
 \frac{\partial \tilde{f}_{\rm nuc}}{\partial n_{B}}
 \right|_{T,\mu_{Q},n_{B} = n_{B}^{(\pi)}}
\end{eqnarray}
As a result, the charge densities
\begin{equation}
 n_{Q}^{(\pi)} = n_{Q}(T,n_{B}^{(\pi)},\mu_{Q})
 = -  \left. 
 \frac{\partial \tilde{f}_{\rm nuc}}{\partial \mu_{Q}}
 \right|_{T,n_{B}^{(\pi)}}
\end{equation}
and entropy densities
\begin{equation}
 s^{(\pi)} = s(T,n_{B}^{(\pi)},\mu_{Q})
  = -  \left. 
 \frac{\partial \tilde{f}_{\rm nuc}}{\partial T}
 \right|_{\mu_{Q},n_{B}^{(\pi)}}
\end{equation}
of the two coexisting phases $\pi = 1,2$ are obtained.
In figure \ref{fig:bino_T10} the phase coexistence boundary and lines of constant
charge chemical potential $\mu_{Q}$ are shown for $T=10$~MeV using the gRDF model
of nuclear matter. The points, where a line of constant $\mu_{Q}$
crosses the binodal, define the baryon densities, $n_{B}^{(1)}$ and
$n_{B}^{(2)}$, and asymmetries of
the coexisting phases.
For all baryonic densities $n_{B}$ inside the coexistence
region, i.e.\
$n_{B}^{(1)} \leq n_{B} \leq n_{B}^{(2)}$, one has
the corresponding charge density
\begin{equation}
 n_{Q}(T,n_{B},\mu_{Q}) = x_{1} n_{Q}^{(1)}
 + x_{2} n_{Q}^{(2)}
\end{equation}
with
\begin{equation}
 x_{1} = \frac{n_{B}^{(2)}-n_{B}}{n_{B}^{(2)}-n_{B}^{(1)}}
\end{equation}
and
\begin{equation}
 x_{2} = \frac{n_{B}-n_{B}^{(1)}}{n_{B}^{(2)}-n_{B}^{(1)}} \: .
\end{equation}
Similarly, the free energy density
\begin{equation}
 f(T,n_{B},n_{Q}) = x_{1} f^{(1)} + x_{2} f^{(2)}
\end{equation}
with
\begin{equation}
 f^{(\pi)} = p^{(\rm coex)} + \mu_{B}^{(\rm coex)} n_{B}^{(\pi)}
 + \mu_{Q}^{(\rm coex)} n_{Q}^{(\pi)}
\end{equation}
and other extensive-like thermodynamic 
quantities are calculated by a linear interpolation.

\subsection{Stellar matter}

In this case, there is an additional conserved charge, the
total (electronic) lepton number. The free energy density
$f(T,n_{B},n_{Q},n_{L})$ depends on the temperature and three
independent densities with three corresponding 
chemical potentials in general. However, the charge neutrality condition requires
$n_{Q}=0$. Instead of the modified free energy density
$\tilde{f}_{\rm nuc}(T,n_{B},\mu_{Q})$ as defined in equation
(\ref{eq:def_ftilde}) it is advantageous to introduce the modified
free energy density
\begin{equation}
 \tilde{f}_{\rm st}(T,n_{B},n_{Q},\mu_{L}) = f(T,n_{B},n_{Q},n_{L}) - \mu_{L} n_{L}
\end{equation}
of stellar matter. Setting $n_{Q}=0$, hence considering a submanifold
in the full parameter space, the construction of the phase transition
can follow the same lines as in the case of nuclear matter by
replacing $n_{Q}$ and $\mu_{Q}$ in the previous subsection by
$n_{L}$ and $\mu_{L}$.

\section{Medium dependent mass shifts of composite particles}
\label{sec:mshifts}

The effective mass 
\begin{equation}
 m_{i}^{(\rm eff)} = N_{i}m_{n} + Z_{i}m_{p} - B(N,Z) - \Gamma_{i\sigma}
 A_{\sigma} + \Delta m_{i}
\end{equation}
of a composite particle $i=(N,Z)$ in dense matter depends on the
interaction with the $\sigma$ meson field and the in-medium mass shift 
\begin{equation}
 \Delta m_{i} = \Delta E_{i}^{(\rm strong)} + \Delta E_{i}^{(\rm Coul)}
\end{equation}
where we consider two contributions. 

The strong mass shift
$\Delta E_{i}^{(\rm strong)}$ includes the effect of the Pauli
exclusion principle causing a blocking of nucleon states in the medium
and the binding energy shift of nuclei due to the strong
interaction. It is parametrized as a function of the temperature $T$
and the effective density
\begin{equation}
 n_{i}^{(\rm eff)} = \frac{2}{A_{i}} \left( N_{i} n_{n}^{(\rm tot)} +
   Z_{i} n_{p}^{(\rm tot)}\right)
\end{equation}
with the total neutron and proton densities, $n_{n}^{(\rm tot)}$ and
$n_{p}^{(\rm tot)}$, counting both nucleons that are free and bound inside
clusters.
In Ref.\ \cite{Typ10} these densities were replaced by approximate values
derived from the strengths of the $\omega$ and $\rho$ meson fields
resulting in a different form of the meson field equations and the
rearrangement terms in the potentials. In the present approach the correct total proton
and neutron densities are used. For light clusters (bound states and
effective two-nucleon resonance states) we assume a product form
\begin{equation}
\label{eq:DE_light}
 \Delta E_{i}^{(\rm strong)}(T,n_{i}^{(\rm eff)}) = 
 f_{i}(n_{i}^{(\rm eff)}) \delta E_{i}^{(\rm Pauli)}(T)
\end{equation}
where $\delta E_{i}^{(\rm Pauli)}(T)$ is given by equations (26) and
(27) of Ref.\ \cite{Typ10} with $n=0$ for two-nucleon and three/four-nucleon
clusters, respectively. See also Refs.\ \cite{Roe09,Roe11} for an improved
parametrization. The prefactor in 
(\ref{eq:DE_light}) is given by the quadratic function
\begin{equation}
 f_{i}(n_{i}^{(\rm eff)}) = n_{i}^{(\rm eff)} \left[ 1 +
   \frac{n_{i}^{(\rm eff)}}{2n_{i}^{(0)}(T)} \right]
\end{equation} 
with the reference density
\begin{equation}
 n_{i}^{(0)}(T) = \frac{B(N_{i},Z_{i})}{\delta E_{i}^{(\rm Pauli)}(T)}
\end{equation}
as in Ref.\ \cite{Typ10}.
For heavy nuclei with $A>4$ we use the pole form
\begin{equation}
\label{eq:DE_heavy}
 \Delta E_{i}^{(\rm strong)}(T,n_{i}^{(\rm eff)}) = 
 \frac{B(N_{i},Z_{i})}{1-x_{i}}
\end{equation}
with the parameter
\begin{equation}
 x_{i}= \frac{n_{i}^{(\rm eff)}}{n_{i}^{(0)}}
\end{equation}
for $x_{i} < 1$. It
depends on the density scale 
\begin{equation}
 n_{i}^{(0)} = \frac{n_{\rm sat}}{1+76/A_{i}}
\end{equation}
with the saturation density $n_{\rm sat}$ of the DD2
parametrization.
For $x_{i} \geq 1$ the particle $i$ is no longer considered to exist
in the medium.

The Coulomb contribution to the mass shift is taken from 
the Wigner-Seitz approximation
\begin{equation}
\label{eq:DE_Coul}
 \Delta E_{i}^{(\rm Coul)} = E_{i}^{(\rm Coul)}
 \left[ - \frac{3}{2} \frac{R_{i}}{R_{i}^{(e)}} 
 +  \frac{1}{2} \left(\frac{R_{i}}{R_{i}^{(e)}}\right)^{3} \right]
\end{equation}
with the Coulomb energy (\ref{eq:E_Coul}) and
the electronic radius 
\begin{equation}
 R_{i}^{(e)} = \left( \frac{3 Z_{i}}{4\pi n_{e}}\right)^{1/3}
\end{equation}
that contains the electron density $n_{e}$ which is assumed to be 
spatially uniform in the present description of dense matter.

\section{Effective degeneracy factors and density of states of nuclei}
\label{sec:geff}

In a medium of finite temperature, not only the ground state of a
nucleus can be populated but also excited states. As a consequence, 
there is a mixture of the nucleus in different excitation states in
warm dense matter. 
The relative probabilities of the different excitation 
states can be found by applying the
appropriate Boltzmann factors depending on the excitation energy $\varepsilon$.
The effect can be summarized by introducing a temperature dependent
degeneracy factor
\begin{eqnarray}
\label{eq:g_T}
 g_{(N,Z)}(T) & = & g_{(N.Z)}^{(gs)} 
 \\ \nonumber & & 
 + \int_{0}^{E_{N,Z}^{(\rm max)}} d\varepsilon \:
 \varrho_{N,Z}^{(\rm exc)}(\varepsilon) \exp\left(-\frac{\varepsilon}{T}\right)
\end{eqnarray}
of a nucleus $(N,Z)$ 
with the degeneracy of ground state $g_{(N,Z)}^{(gs)}=2J_{N,Z}^{(gs)}+1$
and a contribution of excitated states containing the density of
excited states $ \varrho_{N,Z}^{(\rm exc)}(\varepsilon)$.

For the ground state spins $J_{N,Z}^{(gs)}$ experimental values are used
as far as available. They are tabulated in the NUBASE2012 evaluation
\cite{NUBASE2012}. Otherwise we
assume $J_{N,Z}^{(gs)}=0$ for even-even nuclei,
$J_{N,Z}^{(gs)}=1$ for odd-odd nuclei,
and  $J_{N,Z}^{(gs)}=1/2$ for the remaining nuclei.

Following Ref.\ \cite{Gro85}, the density of excited states is assumed to have the form
\begin{eqnarray}
\label{eq:dos}
 \lefteqn{\varrho_{N,Z}^{(\rm exc)}(\varepsilon)}
 \\ \nonumber & = &
 \frac{\sqrt{\pi}}{12}
 \left( \frac{a_{N,Z}^{2}}{4a_{N,Z}^{(n)}a_{N,Z}^{(p)}}\right)^{1/2}
 \frac{\exp\left( \beta_{N,Z} \varepsilon + \frac{a_{N,Z}}{\beta_{N,Z}}\right)}{\left(
 \beta_{N,Z}\varepsilon^{3}\right)^{1/2}}
 \\ \nonumber & & \times
 \frac{1-\exp\left( -\frac{a_{N,Z}}{\beta_{N,Z}}\right)}{\left[ 1 - \frac{1}{2}
 \beta_{N,Z}\varepsilon \exp\left( -
   \frac{a_{N,Z}}{\beta_{N,Z}}\right)\right]^{1/2}} \: .
\end{eqnarray}
Proper values for the function
$\beta_{N,Z}(\varepsilon)$ are obtained by solving the equation
\begin{equation}
 \left( \frac{a_{N,Z}}{\beta_{N,Z}} \right)^{2} =
 a_{N,Z} \varepsilon 
 \left[ 1-\exp\left( -\frac{a_{N,Z}}{\beta_{N,Z}}\right) \right] \: .
\end{equation}
that contains the level density parameter 
\begin{equation}
 a_{N,Z} = a_{N,Z}^{(n)} + a_{N,Z}^{(p)} 
\end{equation}
of a nucleus $(N,Z)$.
For the level density parameters  
of the nucleons
\begin{equation}
 a_{N,Z}^{(i)}  =  \frac{g_{i}}{2}
 \varrho_{N,Z}^{(i)} 
 \frac{\pi^{2}}{3}
\end{equation}
the free Fermi gas estimate
\begin{equation}
 \varrho_{N,Z}^{(i)} 
 = \frac{m_{i}k_{N,Z}^{(i)}V_{N,Z}}{2\pi^{2}}
\end{equation}
is adopted 
with the nuclear volume 
\begin{equation}
 V_{N,Z} = \frac{4\pi}{3} r_{0}^{3} (N+Z)
\end{equation}
(using $r_{0} = 1.4$~fm) and Fermi momenta
\begin{eqnarray}
 k_{N,Z}^{(n)} & = & \left( \frac{N}{V_{N,Z}} \frac{6\pi^{2}}{g_{n}} \right)^{1/3}
 \\
 k_{N,Z}^{(p)} & = & \left( \frac{Z}{V_{N,Z}} \frac{6\pi^{2}}{g_{p}} \right)^{1/3}
\end{eqnarray}
at the neutron and proton Fermi energies.

The advantage of the form (\ref{eq:dos}) is that the level density does not diverge for
$\varepsilon \to 0$ but it remains finite with
\begin{equation}
 \lim_{\varepsilon \to 0}\varrho_{N,Z}^{(\rm exc)}(\varepsilon) =
 \frac{\sqrt{2\pi}e}{12}
 a_{N,Z}   \left(
   \frac{a_{N,Z}^{2}}{4a_{N,Z}^{(n)}a_{N,Z}^{(p)}}\right)^{1/2} 
\end{equation}
in contrast to many Fermi gas models following the 
original ideas of Bethe \cite{Bet36b,Bet37}. This is due to an explicit
separation of the ground state
contribution from the excited-states contribution in deriving the
density of states by an inverse Laplace transformation. 
For high
excitation energies, one finds
$\beta_{N,Z} \to \sqrt{\varepsilon/a_{N,Z}}$ and the usual form
\begin{equation}
 \varrho_{N,Z}^{(\rm exc)}(\varepsilon) \to
 \frac{\sqrt{\pi}}{12}
 \left( \frac{a_{N,Z}^{2}}{4a_{N,Z}^{(n)}a_{N,Z}^{(p)}}\right)^{1/2}
 \frac{\exp\left( 2 \sqrt{a_{N,Z} \varepsilon}\right)}{a_{N,Z}^{1/4} \varepsilon^{5/4}}
\end{equation}
of the Fermi gas model.

Because the density of states (\ref{eq:dos}) is derived in a
low-temperature approximation, 
$\varrho_{N,Z}^{(\rm  exc)}(\varepsilon)$ 
is multiplied with an exponential
damping factor $\exp\left( - \varepsilon/T_{0}\right)$ 
as in Ref.\ \cite{Rad09}. For the parameter $T_{0}$ we use the
critical temperature $T_{\rm crit}$ of the liquid-gas phase transition
in symmetric nuclear matter of the DD2 parametrization.
The maximum excitation energy in (\ref{eq:g_T}) is chosen as
$E_{N,Z}^{(\rm max)}=3(NS_{N,Z}^{(n)}+ZS_{N,Z}^{(p)})/5$ with the neutron and proton
separation energies, $S_{N,Z}^{(n)}$ and $S_{N,Z}^{(p)}$, of the nucleus $(N,Z)$.
For alternative approaches to treat the density of states for nuclei
see Ref.\ \cite{Hem11}.

%

\end{document}